\documentclass[english,structabstract]{aa}
\usepackage{mathptmx}
\usepackage[T1]{fontenc}
\usepackage[latin9]{inputenc}
\usepackage{babel}
\usepackage{array}
\usepackage{textcomp}
\usepackage{multirow}
\usepackage{amsmath}
\usepackage{amssymb}
\usepackage[pdftex]{graphicx} 
\usepackage[authoryear]{natbib}
\usepackage[unicode=true,pdfusetitle,
 bookmarks=true,bookmarksnumbered=false,bookmarksopen=false,
 breaklinks=false,pdfborder={0 0 1},backref=false,colorlinks=false]
 {hyperref}

\makeatletter

\newcommand{\lyxmathsym}[1]{\ifmmode\begingroup\def\b@ld{bold}
  \text{\ifx\math@version\b@ld\bfseries\fi#1}\endgroup\else#1\fi}

\providecommand{\tabularnewline}{\\}

\numberwithin{equation}{section}
\numberwithin{figure}{section}
\numberwithin{table}{section}
\newenvironment{lyxlist}[1]
{\begin{list}{}
{\settowidth{\labelwidth}{#1}
 \setlength{\leftmargin}{\labelwidth}
 \addtolength{\leftmargin}{\labelsep}
 }}
{\end{list}}

\@ifundefined{rotatebox}{\usepackage{graphicx}}{}
\bibpunct{(}{)}{;}{a}{}{,} 
\makeatother

\begin{document}

\title{Gaia Universe Model Snapshot}

\subtitle{A statistical analysis of the expected contents of the Gaia catalogue}

\author{A.C. Robin\inst{1} \and X. Luri\inst{2}  \and C. Reyl\'e\inst{1} \and Y. Isasi\inst{2} \and E. Grux\inst{1} \and S. Blanco-Cuaresma\inst{2} \and F. Arenou\inst{3} \and C. Babusiaux\inst{3} \and M. Belcheva\inst{4} \and R. Drimmel\inst{5}
\and C. Jordi\inst{2}  \and A. Krone-Martins\inst{7} \and E. Masana\inst{2} \and J.C. Mauduit\inst{8} \and F. Mignard\inst{8}\and N. Mowlavi\inst{6} \and B. Rocca-Volmerange\inst{9} \and P. Sartoretti\inst{3} \and E. Slezak\inst{8} \and A. Sozzetti\inst{5}}

\institute{Institut Utinam, CNRS UMR6213, Universit\'e de Franche-Comt\'e, Observatoire de Besan\c{c}on, Besan\c{c}on, France  \\
\email{annie.robin@obs-besancon.fr}
\and Dept. Astronomia i Meteorologia ICCUB-IEEC, Mart\'{\i} i Franqu\`es 1, Barcelona, Spain  \\ 
\email{xluri@am.ub.es}
\and GEPI, Observatoire de Paris, CNRS, Universit\'e Paris Diderot ; 5 Place Jules Janssen, 92190 Meudon, France
\and Department of Astrophysics Astronomy \& Mechanics, Faculty of Physics, University of Athens, Athens, Greece
\and OAT, Torino, Italy  
\and Observatoire de Gen\`{e}ve, Sauverny, Switzerland 
\and SIM, Faculdade de Ci\^{e}ncias, Universidade de Lisboa, Portugal 
\and Observatoire de la C\^{o}te d'Azur, Nice, France  
\and Institut d'Astrophysique de Paris, France \\
}

\offprints{A.C. Robin}

\date{Received ...; Accepted...}

\date{}

\abstract{This study has been developed in the framework of the computational
simulations executed for the preparation of the ESA Gaia astrometric
mission.} {We focus on describing the objects and
characteristics that Gaia will potentially observe without taking
into consideration instrumental effects (detection efficiency, observing errors).} {The theoretical Universe
Model prepared for the Gaia simulation has been statistically analyzed
at a given time. Ingredients of the model are described, giving most attention to the stellar content, 
the double and multiple stars, and variability. } {In this simulation the errors have not been included 
yet. Hence we estimate the number of objects and their theoretical photometric, astrometric and spectroscopic 
characteristics in the case that they are perfectly detected. We show that Gaia will be able to potentially observe
1.1 billion of stars (single or part of multiple star
systems) of which about 2\% are variable stars, 3\% have one or two
exoplanets. At the extragalactic level, observations will be potentially
composed by several millions of galaxies, half million to 1 million of quasars and about 50,000 supernovas
that will occur during the 5 years of mission. The simulated catalogue will be made publicly available 
by the DPAC on the Gaia portal of the ESA web site \url{http://www.rssd.esa.int/gaia/}}
{}

\maketitle

\section{Introduction}

The ESA Gaia astrometric mission has been designed for solving one
of the most difficult yet deeply fundamental challenges in modern
astronomy: to create an extraordinarily precise 3D map of about a
billion stars throughout our Galaxy and beyond \citep{Perryman2001}.

The survey aims to reach completeness at $V_{\text{lim}}\sim20-25$ mag depending
on the color of the object, with astrometric accuracies of about 10$\mu$as
at V=15. In the process, it will map stellar motions and provide
detailed physical properties of each star observed: characterizing
their luminosity, temperature, gravity and elemental composition.
Additionally, it will perform the detection and orbital classification
of tens of thousands of extra-solar planetary systems, and a comprehensive
survey of some $10^{5}-10^{6}$ minor bodies in our solar system,
as well as galaxies in the nearby Universe and distant quasars. 

This massive stellar census will provide the basic observational data
to tackle  important questions related to the origin,
structure and evolutionary history of our Galaxy and new tests of
general relativity and cosmology.

It is clear that the Gaia data analysis will be an enormous task in terms
of expertise, effort and dedicated computing power. In that sense,
the Data Processing and Analysis Consortium (DPAC) is a large pan-European
team of expert scientists and software developers which are officially
responsible for Gaia data processing and analysis%
\footnote{http://www.rssd.esa.int/gaia/dpac%
}. The consortium is divided into specialized units with a unique set
of processing tasks known as Coordination Units (CUs). More precisely,
the CU2 is the task force in charge of the simulation needs for the
work of other CUs, ensuring that reliable data simulations are available
for the various stages of the data processing and analysis development. 
One key piece of the simulation software developed by CU2 for Gaia is the Universe Model that generates the astronomical sources. 

The main goal of the present paper is to analyze the content of a full sky
snapshot (for a given moment in time) of the Universe Model. With
that objective in mind, the article has been organized in two main
parts: in section \ref{sec:Gaia-simulator}, the principal components
of the Gaia simulator are exposed, while the results of the analysis
are detailed in section \ref{sec:Gaia-Universe-Model-Snapshot}, complemented
with a wide variety of diagrams and charts for better understanding.

\begin{figure}
\label{bands}
\begin{centering}
\includegraphics[scale=0.35]{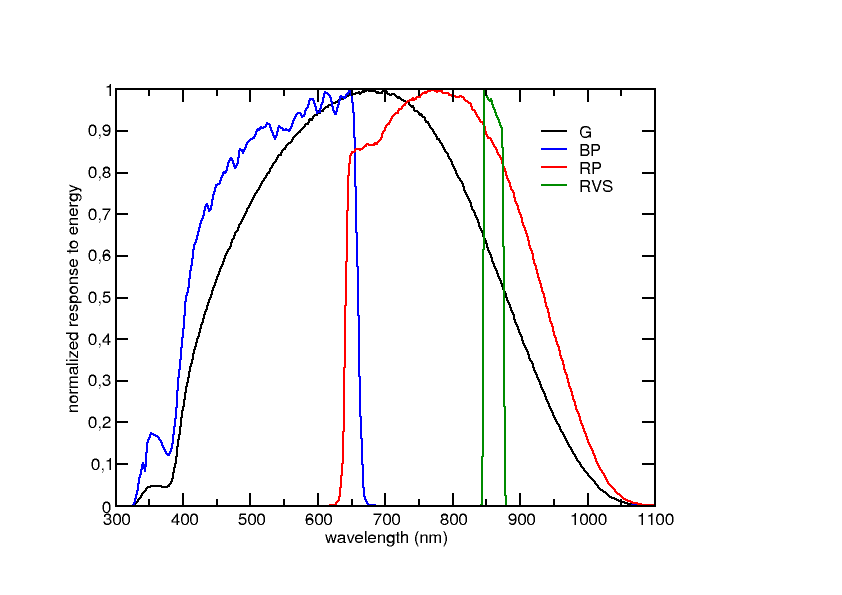}
\par\end{centering}

\caption{Gaia $G$, $G_\mathrm{BP}$ , $G_\mathrm{RP}$ and $G_\mathrm{RVS}$ passbands}
\end{figure}

In order to understand those results, it is important to remark that
four passbands (and their corresponding magnitudes) are associated
with the Gaia instruments: $G$, $G_\mathrm{BP}$ , $G_\mathrm{RP}$ and $G_\mathrm\mathrm{RVS}$ (see fig.~\ref{bands}).

As described in \citet{Jordi2010}, from the astrometric measurements
of unfiltered (white) light, Gaia will yield $G$ magnitudes in a
very broad band covering the $350\lyxmathsym{\textendash}1000$ nm
wavelength range. The $G$ band can be related to Johnson-Cousins
V and I$_{C}$ passband by the  following approximation \citep{Jordi2010} for unreddened stars :


\begin{multline}
 G = V - 0.0257 - 0.0924\;(V-I_C) - 0.1623\;(V-I_C)^2 + \\
     + 0.0090\;(V-I_C)^3
 \end{multline}

for $-0.4<V-I\lesssim6$ . The approximation can be simplifed to $G-V=0.0\pm0.1$
for the range $-0.4<V-I<1.4$. 

Besides, the spectrophotometric instrument \citep{Jordi2010}
consists on two low-resolution slitless spectrographs named 'Blue
Photometer' (BP) and 'Red Photometer' (RP). They cover the wavelength
intervals $330\lyxmathsym{\textendash}680$ nm and $650\lyxmathsym{\textendash}1050$ nm,
and its total flux will yield $G_\mathrm{BP}$ and $G_\mathrm{RP}$ magnitudes.
Additionally, for the brighter stars, Gaia also features a high-resolution
($R=11500$) integral field spectrograph in the range $847\lyxmathsym{\textendash}874$ nm
around CaII triplet named Radial Velocity Spectrometer or RVS instrument.
Its integrated flux will provide the $G_\mathrm{RVS}$ magnitude. 

The RVS will provide the radial velocities of stars up to $G_\mathrm{RVS}=17$
with precisions ranging from 15 km~s$^{-1}$ at the faint end to
1 km~s$^{-1}$ or better at the bright end. Individual abundances
of key chemical elements (e.g. Ca, Mg and Si) will be derived for
stars up to $G_\mathrm{RVS}=12$.

Colour equations to transform Gaia photometric systems to the Johnson, SDSS and Hipparcos photometric systems can be found in \citet{Jordi2010}.


The catalogue described in this paper will be made publicly available by 
the DPAC on the Gaia portal of the ESA web site \url{http://www.rssd.esa.int/gaia/}

\section{\label{sec:Gaia-simulator}Gaia simulator}

Gaia will acquire an enormous quantity of complex and extremely precise
data that will be transmitted daily to a ground station. By the end
of Gaia\textquoteright{}s operational life, around 150 terabytes ($10^{14}$
bytes) will have been transmitted to Earth: some 1000 times the raw
volume from the related Hipparcos mission.  

An extensive and sophisticated Gaia data processing mechanism is required
to yield meaningful results from the collected data. In this sense, an
automatic system has been designed to generate the simulated data
required for the development and testing of the massive data reduction
software.

The Gaia simulator has been organized around a common tool box (named
GaiaSimu library) containing a universe model, an instrument model
and other utilities, such as numerical methods and astronomical tools.
This common tool box is used by several specialized components:
\begin{lyxlist}{00.00.0000}
\item [{GOG~(Gaia~Object~Generator)}] Provides simulations of number
counts and lists of observable objects from the universe model. It
is designed to directly simulate catalog data.
\item [{GIBIS~(Gaia~Instrument~and~Basic~Image~Simulator)}] Simulates
how the Gaia instruments will observe the sky at the pixel level,
using realistic models of the astronomical sources and of the instrument
properties.
\item [{GASS~(GAia~System~Simulator)}] In charge of massive simulations
of the raw telemetry stream generated by Gaia.
\end{lyxlist}
As a component of the GaiaSimu library, the universe model aims at
simulating the characteristics of all the different types of objects
that Gaia will observe: their spatial distribution, photometry, kinematics
and spectra. 

\begin{figure}
\begin{centering}
\includegraphics[scale=0.25]{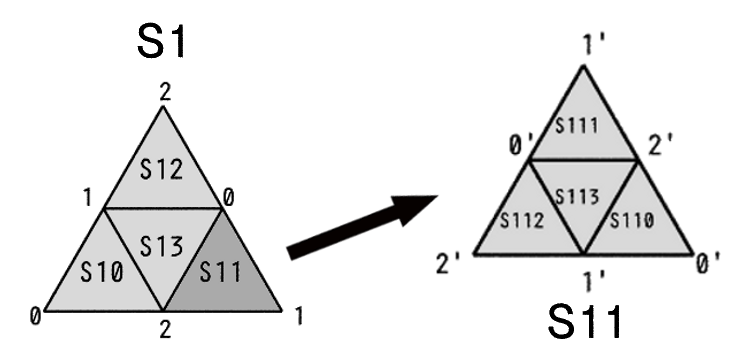}
\par\end{centering}

\caption{\label{fig:Hierarchical-Triangular-Mesh}Hierarchical Triangular Mesh
subdivisions scheme: Each triangle has three vertices labeled 0, 1
and 2. When the area is subdivided in 4 new triangles, the opposite
midpoints are labeled 0', 1' and 2', respectively, and the central
triangle is suffixed with a 3. }
\end{figure}

To handle the simulation, the sky is partitioned by a Hierarchical
Triangular Mesh (HTM) \citep{Kunszt2001}, which subdivides the spherical
surface into triangles in a recursive/multi level process which can
be higher or lower depending on the area density. The first level
divides the northern and southern sphere in four areas each, being
identified by the letter N or S respectively and a number from 0 to
3. The second level for one of these areas (for instance, 'S1') is
generated by subdividing it in four new triangles as stated in figure
\ref{fig:Hierarchical-Triangular-Mesh}. At each level, the area of the different
triangles is almost the same.

The universe model is designed to generate lists of astronomical sources
in given regions of the sky, represented by a concrete HTM area of
a given maximum level. In the present simulation the HTM level is 8, giving a mean size of about 0.1 square degree for each triangular tile.

The distributions of these objects and the statistics of observables
should be as realistic as possible, while the algorithms have been
optimized in order that the simulations can be performed in reasonable
time. Random seeds are fixed to regenerate the same sources at each new generation and in each of the 3 simulators.

The object generation process is divided into three main modules:
\begin{itemize}
\item Solar system object generator such as planets, satellites, asteroids,
comets, etc. This generator is based on a database of known objects
and has not been activated in the present statistical analysis\cite{Tanga2011}.
\item Galactic object generator based on the Besan\c{c}on Galaxy model (BGM
from now on). It creates stellar sources taking into account extinction,
star variability, existence of binary systems and exoplanets.
\item Extragalactic objects generator such as unresolved galaxies, QSO and
supernovas.
\end{itemize}
In the following subsections, the generation of the
different types of objects and the computation of their relevant characteristics are described.

\subsection{Galactic objects}

Galactic objects are generated from a model based on BGM \citep{Robin2003}
which provides the distribution of the stars, their intrinsic parameters
and their motions. The stellar population synthesis combines:
\begin{itemize}
\item Theoretical considerations such as stellar evolution, galactic evolution
and dynamics.
\item Observational facts such as the local luminosity function, the age-velocity
dispersion relation, the age-metallicity relation.
\end{itemize}
The result is a comprehensive description of the stellar components
of the Galaxy with their physical characteristics (e.g. temperature,
mass, gravity, chemical composition and motions).

The Galaxy model is formed by four stellar populations constructed
with different model parameters. For each population the stellar content is defined by the Hess diagram according to the age 
and metallicity characteristics. The populations considered here are:
\begin{itemize}
\item The thin disc: young stars with solar metallicity in the mean. It is additionally
divided in seven isothermal components of ages varying from 0--0.15
Gyr for the youngest to 7--10 Gyr for the oldest. 
\item The thick disc: in terms of metallicity, age and kinematics, stars are intermediate between
the thin disc and the stellar halo.
\item The stellar halo (spheroid): old and metal poor stars.
\item The outer bulge: old stars with metallicities similar to the ones
in the thin disc.
\end{itemize}

The simulations are done using the equation of stellar statistics.
Specifying a direction and a distance $r_\text{max}$, the model generates
a star catalog using the following equation:

\begin{equation}
N=\int_{0}^{r_\text{max}}\sum_{i=1}^{n}\rho_{i}\left(R,\theta,z,Age\right)\times\Phi\left(M_V,T_\mathrm{eff},Age\right)\omega r^{2}dr
\end{equation}
where $\rho\left(R,\theta,z,Age\right)$ is the stellar density
law with galactocentric coordinates $\left(R,\theta,z\right)$, also described in table~\ref{table_density}, $\Phi\left(M_V,T_\mathrm{eff},Age\right)$
is the number of stars per square parsec in a given cell of the HR Diagram
($M_V, T_\mathrm{eff}, $Age) for a given age range near the sun, $\omega$ is
the solid angle and $r$ the distance to the sun.

The functions  $\rho\left(R,\theta,z,Age\right)$ (density laws) and $\Phi\left(M_V,T_\mathrm{eff},Age\right)$
(Hess diagrams) are specific for each population, and established using theoretical  and empirical constraints, as described below.

In a given volume element having an expected density of N, $\approx$ N stars are generated using a Poisson distribution. 
After generating the corresponding number of stars,  each star is assigned its intrinsic
attributes (age, effective temperature, bolometric magnitude, $U,V,W$
velocities, distance) and corresponding observational parameters (apparent magnitudes,
colors, proper motions, radial velocities, etc) and finally affected by the implemented
3D extinction model from \citet{Drimmel2003}.

\subsubsection{Density laws}

\begin{table}
\begin{centering}
\caption{\label{tab:Local-mass-density}Local mass density $\rho_{0}$ for different populations. Axis ratio $\epsilon$ are given as a function of age for the disc population and the spheroid.}
\label{table_density}
\begin{tabular}{|c|c|c|c|}
\hline 
Population & Age (Gyr) & Local density $\left(M_{\odot}\mathrm{pc}^{-3}\right)$ & $\epsilon$ \tabularnewline
\hline 
\hline 
Disc & 0--0.15 & $4.0\times10^{-3}$& 0.0140 \tabularnewline
\cline{2-4} 
 & 0.15--1 & $7.9\times10^{-3}$& 0.0268\tabularnewline
\cline{2-4} 
 & 1--2 & $6.2\times10^{-3}$& 0.0375\tabularnewline
\cline{2-4} 
 & 2--3 & $4.0\times10^{-3}$& 0.0551\tabularnewline
\cline{2-4} 
 & 3--5 & $4.0\times10^{-3}$& 0.0696\tabularnewline
\cline{2-4} 
 & 5--7 & $4.9\times10^{-3}$& 0.0785\tabularnewline
\cline{2-4} 
 & 7--10 & $6.6\times10^{-3}$& 0.0791\tabularnewline
\cline{2-4} 
 & WD & $3.96\times10^{-3}$& - \tabularnewline
\hline 
Thick disc & 11 & $1.34\times10^{-3}$& - \tabularnewline
\cline{2-4} 
 & WD & $3.04\times10^{-4}$& - \tabularnewline
\hline 
Spheroid & 14 & $9.32\times10^{-6}$& 0.76 \tabularnewline
\hline 
\end{tabular}
\par\end{centering}

\end{table}

Density laws allow to extrapolate what is observed in the solar neighbourhood
(i.e. local densities) to the rest of the Galaxy. 
The density law of each population has been described in \citet{Robin2003}.
However the disc and bulge density have been slightly changed. Revised density laws are given in table \ref{tab:Density-laws}.
It is worth noting
that the local density assigned to the seven sub-populations of the
thin disc and its scale height $z$ has been defined by a dynamically
self-consistent process using the Galactic potential and Boltzman
equations \citep{Bienayme1987}.
In this table for simplicity the density formulae do not include the
warp and flare, which are added as a modification of the position
and thickness of the mid-plane. 

The flare is the increase of the thickness of the disc with galactocentric
distance:

\begin{equation}
k_\mathrm{flare}=1.+g_\mathrm{flare}\times\left(R-R_\mathrm{flare}\right)
\end{equation}
with $g_\mathrm{flare}=0.545\times10^{-6} ~\mathrm{pc}^{-1}$ and $R_\mathrm{flare}=9500$ pc.
The warp is modeled as a symmetrical S-shape warp with a linear slope
of 0.09 starting at galactocentric distance of 8400 pc \citep{Reyle2009}.
Moreover a spiral structure has been added with 2 arms, as determined
in a preliminary study by De Amores \& Robin (in prep.). The parameters
of the arms are given in table \ref{tab:Provisional-parameters-of-arms}. 

\begin{table}
\begin{centering}
\caption{\label{tab:Provisional-parameters-of-arms}Parameters of the 2 arms
of the spiral structure.}
\begin{tabular}{|c|c|c|}
\hline 
Parameter & 1st arm & 2nd arm\tabularnewline
\hline 
\hline 
Internal radius in kpc  & 3.426  & 3.426 \tabularnewline
\hline 
Pitch angle in radian  & 4.027  & 3.426 \tabularnewline
\hline 
Phase angle of start in radian  & 0.188  & 2.677 \tabularnewline
\hline 
Amplitude  & 1.823  & 2.013 \tabularnewline
\hline 
Thickness in the plane  & 4.804  & 4.964 \tabularnewline
\hline 
\end{tabular}
\par\end{centering}

\end{table}

\begin{table*}
\begin{centering}
\caption{\label{tab:Density-laws}Density laws where $\rho_{0}$ is the local
mass density, $d_{0}$ a normalization factor to have a density of 1 at the solar position, $k_\mathrm{flare}$ the flare
factor and $a=\sqrt{R^{2}+\left(\frac{z}{\epsilon}\right)^{2}}$ in
kpc with $\epsilon$ being the axis ratio and ($R,z$) the cylindrical galactic coordinates. Local
density $\rho_{0}$ and axial ratio $\epsilon$ can be found in table
\ref{tab:Local-mass-density}. For simplicity the disc density law
is given here without the warp and flare (see text for their characteristics).
For the bulge, $x,y,z$ are in the bulge reference frame and values
of $N,\, x_{0},\, y_{0},\, z_{0},\, R_{c}$ as well as angles from
the main axis to the Galaxy reference frame are given in \citet{Robin2003},
table 5. }
\begin{tabular}{|c|>{\centering}m{0.95\columnwidth}|c|}
\hline 
Population & Density laws & \tabularnewline
\hline 
\hline 
Disc & $\rho_{0}/d_{0}/k_\mathrm{flare}\times\left\{ e^{-\left(\frac{a}{h_{R_{+}}}\right)^{2}}-e^{-\left(\frac{a}{h_{R_{-}}}\right)^{2}}\right\} $ 

with $h_{R_{+}}=5000$ pc and $h_{R_{-}}=3000$ pc & if age $\leq0.15$ Gyr\tabularnewline
 & $\rho_{0}/d_{0}/k_\mathrm{flare}\times\left\{ e^{-\left(0.25+\frac{a^{2}}{h_{R_{+}}^{2}}\right)^{\frac{1}{2}}}-e^{-\left(0.25+\frac{a^{2}}{h_{R_{-}}^{2}}\right)^{\frac{1}{2}}}\right\} $ 

with $h_{R_{+}}=2530$ pc and $h_{R_{-}}=1320$ pc & if age $>0.15$ Gyr \tabularnewline
\hline 
Thick disc & $\rho_{0}/d_{0}/k_\mathrm{flare}\times e^{-\frac{R-R_{\odot}}{h_{R}}}\times\left(1-\frac{1/h_{z}}{x_{l}\times\left(2.+x_{l}/h_{z}\right)}\times z^{2}\right)$  & if $\left|z\right|\leq x_{l},x_{l}=72$ pc\tabularnewline
 & $\rho_{0}\times e^{-\frac{R-R_{\odot}}{h_{R}}}\times\frac{e^{x_{l}/h_{z}}}{1+x_{l}/2h_{z}}e^{-\frac{\left|z\right|}{h_{z}}}$ 

with $h_{R}=4000$ pc and $h_{z}=1200$ pc & if $\left|z\right|>x_{l},x_{l}=72$ pc\tabularnewline
\hline 
Spheroid & $\rho_{0}/d_{0}\times\left(\frac{a_{c}}{R_{\odot}}\right)^{-2.44}$ & if $a\leq a_{c},\, a_{c}=500$ pc\tabularnewline
 & $\rho_{0}\times\left(\frac{a}{R_{\odot}}\right)^{-2.44}$ & if $a>a_{c},\, a_{c}=500$ pc\tabularnewline
\hline 
Bulge & $N\times e^{-0.5\times r_{s}^{2}}$ & $\sqrt{x^{2}+y^{2}}<R_{c}$\tabularnewline
 & $N\times e^{-0.5\times r_{s}^{2}}\times e^{-0.5\left(\frac{\sqrt{x^{2}+y^{2}}}{0.5}\right)^{2}}$

with $r_{s}^{2}=\sqrt{\left[\left(\frac{x}{x_{0}}\right)^{2}+\left(\frac{y}{y_{0}}\right)^{2}\right]^{2}+\left(\frac{z}{z_{0}}\right)^{4}}$ & $\sqrt{x^{2}+y^{2}}>R_{c}$\tabularnewline
\hline 
\end{tabular}
\par\end{centering}

\end{table*}

\subsubsection{Magnitude - Temperature - Age distribution}

The distribution of stars in the HR diagram $\Phi\left(M_V,T_\mathrm{eff}\right)$
is based on the Initial Mass Function (IMF) and Star Formation Rate (SFR) observed
in the solar neighbourhood (see table \ref{tab:IMF-SFR}). For each
population, the SFR determines how much stellar mass is created at
a given formation epoch, while the IMF distributes this mass into
stars of different sizes. Then, the model brings every star created
in each formation epoch to the present day considering evolutionary
tracks and the population age.

For the thin disc, the distribution in the Hess diagram splits into several age bins. It is obtained from an evolutionary model which starts with a mass of gas,
generates stars of different masses assuming an IMF
and a SFR history, and makes these stars evolve along
evolutionary tracks. The evolution model is described in \citet{Haywood1997,Haywood1997a}.
It produces a file describing the distribution
of stars per element volume in the space ($M_V $, T$_\mathrm{eff} $, Age). Similar
Hess diagrams are also produced for the bulge, the thick disc and
the spheroid populations, assuming a single burst of star formation
and ages of 10 Gyr, 11 Gyr and 14 Gyr respectively using \cite{Bergbush} isochrones. 

The stellar luminosity function is the one of primary stars (single
stars, or primary stars in multiple systems) and is normalized to the luminosity
functions in the solar neighbourhood
\citep{Reid2002}.

A summary of age and metallicities, star formation history and IMF 
for each population is given in tables \ref{tab:IMF-SFR}
and \ref{tab:Age-Metal}.

White dwarfs are taken into account using the \cite{Wood} luminosity function for the disc and \cite{Chabrier1999} for the halo. Bulge white dwarfs are not considered. 
Additionally, some rare objects such as Be
stars, peculiar metallicity stars and Wolf Rayet stars have also been
added (see section 2.1.6).

\begin{table}
\noindent \begin{centering}
\caption{\label{tab:IMF-SFR}IMF and SFR for each population for primary stars.}
\begin{tabular}{|c|c|>{\centering}p{3.5cm}|c|}
\hline 
 & Age (Gyr) & IMF & SRF\tabularnewline
\hline 
\hline 
Disc & 0-10 & $\begin{alignedat}{1}f\left(m\right)=\frac{dn}{dm}\propto m^{-\alpha}\\
\alpha=1.1,   0.07<m<0.6 M_{\odot}\\
\alpha=1.6,   0.6<m<1 M_{\odot}\\
\alpha=3.0,    m>1 M_{\odot}
\end{alignedat}
$ & constant\tabularnewline
\hline 
Thick disc & 11 & $f\left(m\right)=\frac{dn}{dm}\propto m^{-0.5}$ & one burst\tabularnewline
\hline 
Stellar halo & 14 & $f\left(m\right)=\frac{dn}{dm}\propto m^{-0.5}$ & one burst\tabularnewline
\hline 
Bulge & 10 & $f\left(m\right)=\frac{dn}{dm}\propto m^{-2.35}$ 

for $m>0.7M_{\odot}$ & one burst\tabularnewline
\hline 
\end{tabular}
\par\end{centering}

\end{table}

\subsubsection{\label{sub:Metallicity}Metallicity}

Contrarily to \cite{Robin2003} metallicities {[}Fe/H{]} are computed through an empirical age-metallicity
relation $\psi\left(Z,\mathrm{Age}\right)$ from \citet{Haywood2008}. The mean thick disc and spheroid metallicities have also been revised. For each
age and population component the metallicity is drawn from a gaussian,
with a mean and a dispersion as given in table \ref{tab:Age-Metal}.
One also accounts for radial metallicity gradient for the thin disc,
$-0.7$ dex/kpc, but no vertical metallicity gradient.

\begin{table}
\noindent \begin{centering}
\caption{\label{tab:Age-Metal}Age, metallicity and radial metallicity gradients.}
\begin{tabular}{|c|>{\centering}p{0.1\columnwidth}|>{\centering}p{0.3\columnwidth}|>{\centering}p{0.12\columnwidth}|}
\hline 
 & Age (Gyr) & <{[}Fe/H{]}>      (dex) & $\frac{d\left[Fe/H\right]}{dR}$ (dex/kpc)\tabularnewline
\hline 
\hline 
Disc & 0--0.15

0.15--1

1--2

2--3

3--5

5--7

7--10 & $0.01\pm0.010$

$0.00\pm0.11$

$-0.02\pm0.12$

$-0.03\pm0.125$

$-0.05\pm0.135$

$-0.09\pm0.16$

$-0.12\pm0.18$ & $-0.07$\tabularnewline
\hline 
Thick disc & 11 & $-0.50\pm0.30$ & $0.00$\tabularnewline
\hline 
Stellar Halo & 14 & $-1.5\pm0.50$ & $0.00$\tabularnewline
\hline 
Bulge & 10 & $0.00\pm0.20$ & $0.00$\tabularnewline
\hline 
\end{tabular}
\par\end{centering}

\end{table}

\subsubsection{Alpha elements - Metallicity relation}

Alpha element abundances are computed as a function of the population and the
metallicity. For the halo, the abundance in
alpha element is drawn from a random around a mean value, while for
the thin disc, thick disc, and bulge populations it depends on {[}Fe/H{]}.
Formulas are given in table \ref{tab:Alpha-Metal}.

\begin{table*}
\noindent \begin{centering}
\caption{\label{tab:Alpha-Metal}Alpha element abundances and metallicity relation estimated from Bensby \& Feltzing (2009) for the thin and thick disc, and Gonzales et al (2011) for the bulge.}
\begin{tabular}{|c|c|c|}
\hline 
 & <{[}$\alpha$/Fe{]}> (dex) & Dispersion\tabularnewline
\hline 
\hline 
Disc & $0.01043-0.13\times\left[Fe/H\right] + 0.197*[Fe/H]^{2}+0.1882*[Fe/H]^{3}$ & $0.02$\tabularnewline
\hline 
Thick disc & $0.392-e^{1.19375\times\left[Fe/H\right]-1.3038}$ & $0.05$\tabularnewline
\hline 
Stellar Halo & $0.4$ & $0.05$\tabularnewline
\hline 
Bulge & $-0.334\times\left[Fe/H\right]+0.134$ & $0.05$\tabularnewline
\hline 
\end{tabular}
\par\end{centering}

\end{table*}

\subsubsection{Age - velocity dispersion}

Age-velocity dispersion relation is obtained from \citet{Gomez1997}
for the thin disc, while \citet{Ojha1995,Ojha1999} determined the
velocity ellipsoid of the thick disc, which has been used in the model
(see table \ref{tab:Velocity-dispersion}).

\begin{table}
\begin{centering}
\caption{\label{tab:Velocity-dispersion}Velocity dispersions, asymmetric drift
$V_{ad}$ at the solar position and velocity dispersion gradient.
W axis is pointing the north galactic pole, U the galactic center
and V is tangential to the rotational motion. It is worth noting that
$\frac{d\ln\left(\sigma_{U}^{2}\right)}{dR}=0.2$ for the disc.}
\begin{tabular}{|c|>{\centering}p{1cm}|>{\centering}p{1cm}|>{\centering}p{1cm}|>{\centering}p{1cm}|>{\centering}p{1cm}|}
\hline 
 & Age

(Gyr) & $\sigma_{U}$

$\left(km\, s^{-1}\right)$ & $\sigma_{V}$

$\left(km\, s^{-1}\right)$ & $\sigma_{W}$

$\left(km\, s^{-1}\right)$ & $V_{ad}$

$\left(km\, s^{-1}\right)$\tabularnewline
\hline 
\hline 
Disc & 0--0.15 & 16.7  & 10.8  & 6 & 3.5 \tabularnewline
\cline{2-6} 
 & 0.15--1 & 19.8  & 12.8  & 8 & 3.1 \tabularnewline
\cline{2-6} 
 & 1--2 & 27.2  & 17.6  & 10 & 5.8 \tabularnewline
\cline{2-6} 
 & 2--3 & 30.2  & 19.5  & 13.2  & 7.3 \tabularnewline
\cline{2-6} 
 & 3--5 & 36.7  & 23.7  & 15.8  & 10.8 \tabularnewline
\cline{2-6} 
 & 5--7 & 43.1  & 27.8  & 17.4  & 14.8 \tabularnewline
\cline{2-6} 
 & 7--10 & 43.1  & 27.8  & 17.5  & 14.8 \tabularnewline
\hline 
Thick disc & 11 & 67  & 51  & 42  & 53 \tabularnewline
\hline 
Spheroid & 14 & 131  & 106  & 85 & 226 \tabularnewline
\hline 
Bulge & 10 & 113  & 115  & 100  & 79\tabularnewline
\hline 
\end{tabular}
\par\end{centering}

\end{table}

\subsubsection{Stellar rotation}

The rotation of each star is simulated following specifications from \citet{Cox2000}. 
The rotation velocity is computed as a function of luminosity and spectral type. 
Then $vsini$ is computed for random values of the inclination of the star's rotation axis.

\subsubsection{\label{sub:Rare-objects}Rare objects}

For the needs of the simulator, some rare objects have been added
to the BGM Hess diagram:

Be stars: this is a transient state of B-type stars with a gaseous
disc that is formed of material ejected from the star (Be stars are
typically variable). Prominent emission lines of hydrogen are found
in its spectrum because of re-processing stellar ultraviolet light
in the gaseous disc. Additionally, infrared excess and polarization
is detected as a result from the scattering of stellar light in the
disc.\\
\\
Oe and Be stars are simulated as a proportion of 29\% of O7-B4 stars, 20\%
of B5-B7 stars and 3\% of B8-B9 stars \citep{Jaschek1988,Zorec1997}.
For these objects, the ratio between the envelope radius and the stellar
radius is linked with the line strength in order to be able to determine
their spectrum. Over the time, this ratio changes between 1.2 and
6.9 to simulate the variation of the emission lines.

Two types of peculiar metallicity stars are simulated, following \citet{Kurtz1982,Kochukhov2007,Stift2009},
for A and B stars that have a much slower rotation than normal:

\begin{itemize}
\item Am stars have strong and often variable absorption lines of metals
such as zinc, strontium, zirconium, and barium, and deficiencies of
others, such as calcium and scandium. These anomalies with respect to A-type
stars are due to the fact that the elements that absorb more light
are pushed towards the surface, while others sink under the force
of gravity.\\
\\
In the model, 12\% of A stars on the main sequence in the $T_\mathrm{eff}$
range 7400 K - 10200 K are set to be Am stars. 70\% of these Am stars
belong to a binary system with period from 2.5 to 100 days. They are
forced to be slow rotators: 67\% have a projection of rotation velocity
$v_\mathrm{sin\, i}<50$ km/s and 23\% have $50<v_\mathrm{sin\, i}<100$ km/s.\\

\item Ap/Bp stars present overabundances of some metals, such as strontium,
chromium, europium, praseodymium and neodymium, which might be connected
to the present stronger magnetic fields than classical A or B type
star.\\
\\
In the simulation, 8\% of A and B stars on the main sequence in the
$T_\mathrm{eff}$ range 8000 K -- 15000 K are set to be Ap or Bp stars. They
are forced to have a smaller rotation : $v_\mathrm{sin\, i}<120$~km/s. \\

\end{itemize}

Wolf Rayet (WR) stars are hot and massive stars with a high rate of
mass loss by means of a very strong stellar wind. There are 3 classes
based on their spectra: the WN stars (nitrogen dominant, some carbon),
WC stars (carbon dominant, no nitrogen) and the rare WO stars with
C/O < 1. \\
\\
WR densities are computed following the observed statistics from the
VIIth catalogue of Galactic Wolf-Rayet stars of \citet{vanDerHucht2001}.
The local column density of WR stars is $2.9\times10^{\lyxmathsym{\textminus}6}\mathrm{pc}^{\text{\textminus}2}$
or volume density $2.37\times10^{\lyxmathsym{\textminus}8}\mathrm{pc}^{\text{\textminus}3}$.
Among them, 50\% are WN, 46\% are WC and 3.6\% are WO. The absolute
magnitude, colors, effective temperature, gravity and mass have been
estimated from the literature. The masses and effective temperature
vary considerably from one author to another. As a conservative value,
it is assumed that the WR stars have masses of 10 $M_{\odot}$ in
the mean, an absolute V magnitude of -4, a gravity of -0.5,
and an effective temperature of 50,000 K.

\subsubsection{\label{sub:Binary-systems}Binary systems}

The BGM model produces only single stars which densities have been normalized to
follow the luminosity function (LF) of single stars and primaries
in the solar neighbourhood \citep{Reid2002}. The
IMF, inferred from the LF and the mass-luminosity relation, goes
down to the hydrogen burning limit, and include disc stars down to
$M_V=24$. It corresponds to spectral types down to about L5.

In the Gaia simulation multiple star systems are generated with some
probability \citep{2011AIPC.1346..107A} 
increasing with the mass of
the primary star obtained from the BGM model.

The mass of the companion is then obtained through a given statistical
relation $q=\frac{M_{2}}{M_{1}}=f\left(M_{1}\right)$
which  depends on period and mass ranges, ensuring that
the total number of stars and their distribution is compatible with
the statistical observations and checking that the pairing is realistic. The mass and age of secondary determine physical parameters computed using the Hess diagram distribution in the Besan\c{c}on model. Although, for the case of PMS stars, it appears that pairing has been done is some cases with main sequence stars due to the resolution in age which is not good enough to distinguish them. It will be improved in further simulations.

It is worth noting that while, observationally, the primary of a system
is conventionally the brighter, the model uses here the other convention,
i.e. the primary is the one with the largest mass, and consequently
the generated mass ratio is constrained to be $0<q\leq1$. 

The separation of the components (AU) is chosen with a Gaussian probability
with different mean values depending on the stars' masses  
(a smaller average separation for low mass stars). Through Kepler's third law, the
separation and masses then give the orbital period. 
The average orbital eccentricity (a perfect circular orbit corresponds
to $e=0$) depends on the period by the following relation:

\begin{equation}
E\left[e\right]=a\left(b-e^{-c\log\left(P\right)}\right)
\end{equation}
where $a,b,c$ are constant with different values depending on the
star's spectral type (values can be found in table 1 of \citet{Arenou2010})

To describe the orientation of the orbit, three angles are chosen
randomly:
\begin{itemize}
\item The argument of the periastron $\omega_{2}$ uniformly in $\left[0,\,2\pi\right[$, 
\item The position angle of the node $\Omega$ uniformly in $\left[0,\,2\pi\right[$ 
\item The inclination $i$ uniformly random in $\cos\left(i\right)$. 
\end{itemize}
The moment at which stars are closest together (the periastron date
$T$) is also chosen uniformly between 0 and the period $P$. 

A large effort has been put at trying to ensure that simulated multiple
stars numbers are in accordance with latest fractions known from available
observations. A more detailed description can be found 
in \citet{2011AIPC.1346..107A}.

Although the present paper describes the content
of the Universe Model at a fixed moment in time, it should be reminded
that the model is being used to simulate the Gaia observations. Thus, obviously,
the astrometric, photometric and spectroscopic observables of a multiple system 
vary in time according to the orbital properties. This means
that e.g. the apparent path of the photocentre of an 
unresolved binary will reflect the orbital 
motion through positional and radial velocity changes, 
or that the light curve of an eclipsing binary will vary in each band.

\subsubsection{Variable stars}
\begin{itemize}
\item Regular and semi-regular variables
\end{itemize}
Depending on their position in the HR Diagram, the generated stars
have a probability of being one of the six types of regular and semi-regular
variable stars considered in the simulator:
\begin{description}
\item [{Cepheids:}] Supergiant stars which undergo pulsations with very
regular periods on the order of days to months. Their luminosity is
directly related to their period of variation.
\item [{$\delta$Scuti:}] Similar to Cepheids but rather fainter, and with
shorter periods.
\item [{RR~Lyrae:}] Much more common than Cepheids, but also much less
luminous. Their period is shorter, typically less than one day. They
are classified into 

\begin{description}
\item [{RRab:}] Asymmetric light curves (they are the majority type).
\item [{RRc:}] Nearly symmetric light curves (sometimes sinusoidal).
\end{description}
\item [{Gamma~Doradus:}] Display variations in luminosity due to non-radial
pulsations of their surface. Periods of the order of one day.
\item [{RoAp:}] Rapidly oscillating Ap stars are a subtype of the Ap star
class (see section \ref{sub:Rare-objects}) that exhibits short-timescale
rapid photometric or radial velocity variations. Periods on the order
of minutes. 
\item [{ZZceti:}] Pulsating white dwarf with hydrogen atmosphere. These
stars have periods between seconds to minutes.
\item [{Miras:}] Cool red supergiants, which are undergoing very large
pulsations (order of months).
\item [{Semiregular:}] Usually red giants or supergiants that show a definite
period on occasion, but also go through periods of irregular variation.
Periods lie in the range from 20 to more than 2000 days.
\item [{ACV~($\alpha$~Canes~Venaticorum):}] Stars with strong magnetic
fields whose variability is caused by axial rotation with respect
to the observer.
\end{description}
A summary of the variable type characteristics is given in table \ref{tab:Characteristics-of-the-variables} and
the description of their light curve is given in table~\ref{tab:Variables-light-curve}

\begin{table}
\begin{centering}
\caption{\label{tab:Characteristics-of-the-variables}Characteristics of the
variable types. Localization in the (spectral type, luminosity class)
diagram, probability of a star to be variable in this region, stellar population and metallicity
range which are concerned. }
\begin{tabular}{|c|c|c|c|c|c|}
\hline
Type&  Spec.&  Lum.&  Proba&  Pop&  {[}Fe/H{]}\tabularnewline
\hline
\hline
$\delta$ Scuti (a)&  A0:F2&  III&  0.3&  all&  all\tabularnewline
\hline
$\delta$ Scuti (b)&  A1:F3&  IV:V&  0.3&  all&  all\tabularnewline
\hline
ACV (a)&  B5:B9&  V&  0.016&  thin disc&  -1 to 1\tabularnewline
\hline
ACV (b)&  A0:A8&  IV:V&  0.01&  thin disc&  -1 to 1\tabularnewline
\hline
Cepheid&  F5:G0&  I:III&  0.3&  thin disc&  -1 to 1\tabularnewline
\hline
RRab&  A8:F5&  III&  0.4&  spheroid&  all\tabularnewline
\hline
RRc&  A8:F5&  III&  0.1&  spheroid&  all\tabularnewline
\hline
RoAp&  A0:A9&  V&  0.001&  thin disc&  -1 to 1\tabularnewline
\hline
SemiReg (a)&  K5:K9&  III&  0.5&  all&  all\tabularnewline
\hline
SemiReg (b)&  M0:M9&  III&  0.9&  all&  all\tabularnewline
\hline
Miras&  M0:M9&  I:III&  1.0&  all&  all\tabularnewline
\hline
ZZCeti&  White dwarf&  -&  1.0&  all&  all\tabularnewline
\hline
GammaDor&  F0:F5&  V&  0.3&  all&  all\tabularnewline
\hline
\end{tabular}
\par\end{centering}

\end{table}

Period and amplitude are taken randomly from a 2D distribution defined
for each variability type \citep{Eyer2005}. For Cepheids, a period-luminosity
relation is also included log(P)=$(-M_{V}+1.42)/2.78$ \citep{Molinaro2011}. For Miras the relation is log(P)=$(-M_\mathrm{Bol}+2.06)/2.54$ \citep{Feast1989}.
The different light curve models for each variability type
are described in \citet{Reyle2007}. 

The variation of the radius and radial velocity are computed accordingly
to the light variation, for stars with radial pulsations (RRab, RRc,
Cepheids, $\delta$ Scuti, SemiRegular, and Miras).

\begin{table*}
\begin{centering}
\caption{\label{tab:Variables-light-curve}Light curves of the regular or semi-regular
variable stars where $\omega\,_{t}=2\pi\frac{t}{P}+\phi$ ($P$ is the
period, $\phi$ is the phase)}
\begin{tabular}{|c|c|}
\hline 
Type & Light curve\tabularnewline
\hline 
\hline 
Cepheid & $\begin{aligned}S= & 0.148\sin\left(\omega\,_{t}-20.76\right)+0.1419\sin\left(2\omega\,_{t}-63.76\right)\\
 & +0.0664\sin\left(3\omega\,_{t}-91.57\right)+0.0354\sin\left(4\omega\,_{t}-112.62\right)\\
 & +0.020\sin\left(5\omega\,_{t}-129.47\right)
\end{aligned}
$\tabularnewline
\hline 
$\delta$ Scuti, RoAp, RRc, Miras & $S=0.5\sin\left(\omega\,_{t}\right)$\tabularnewline
\hline 
ACV & $S=-0.5\cos\left(2\omega\,_{t}\right)\frac{1-f\times\cos\left(\omega\,_{t}\right)}{1+f/2}$
where $f$ is a random number in [0;1]\tabularnewline
\cline{1-2} 
SemiReg &  Inverse Fourier Transform of a gaussian in frequency space \tabularnewline
\hline 
\end{tabular}
\par\end{centering}

\end{table*}

\begin{itemize}
\item Dwarf and classical novae
\end{itemize}
Dwarf novae and Classical novae are cataclysmic variable stars consisting
of close binary star systems in which one of the components is a white
dwarf that accretes matter from its companion. 

Classical novae result from the fusion and detonation of accreted
hydrogen, while current theory suggests that dwarf novae result from
instability in the accretion disc that leads to releases of large
amounts of gravitational potential energy. Luminosity of dwarf novae
is lower than classical ones and it increases with the recurrence
interval as well as the orbital period.

The model simulates half of the white dwarfs in close binary systems
with period smaller than 14 hours as dwarf novae. The light curve is
simulated by a linear increase followed by an exponential decrease.
The time between two bursts, the amplitude, the rising time and the
decay time are drawn from gaussian distributions derived from OGLE
observations (Wyrzykowski \& Skowron, private communication). 

The other half of white dwarfs in such systems is simulated as a Classical
novae. 
\begin{itemize}
\item M-dwarf flares
\end{itemize}
Flares are due to magnetic reconnection in the stellar atmospheres.
These events can produce dramatic increases in brightness when they
take place in M dwarfs and brown dwarfs.

The statistics used in the model for M-dwarf flares are mainly based
on \citet{Kowalski2009} and their study on SDSS data: 0.1\% of M0-M1
dwarfs, 0.6\% of M2-M3 dwarfs, and 5.6\% of M4-M6 dwarfs are flaring.
The light curve for magnitude $m$ is described as follows: 

\begin{flalign}
f & =1+e^{-\left(t-t_{0}\right)/\tau} & \, t>t_{0}\nonumber \\
f & =1 & \, t<t_{0}\\
m & = m_{0}-1.32877\times A\times2.5\times\log\left(f\right)\nonumber 
\end{flalign}
where $t_{0}$ is the time of the maximum, $\tau$ is the decay time
(in days, random between 1 and 15 minutes), $m_{0}$ is the baseline
magnitude of the source star, $A$ is the amplitude in magnitudes
(drawn from a gaussian gaussian distribution with $\sigma=1$ and
$x_{0}=1.2$).
\begin{itemize}
\item Eclipsing binaries
\end{itemize}
Eclipsing binaries, while being variables, are treated as binaries
and the eclipses are computed from the orbits of the components. See
section \ref{sub:Binary-systems}.
\begin{itemize}
\item Microlensing events
\end{itemize}
Gravitational microlensing is an astronomical phenomenon due to the
gravitational lens effect, where distribution of matter between a
distant source and an observer is capable of bending (lensing) the
light from the source. The magnification effect permits the observation
of faint objects such as brown dwarfs.

In the model, microlensing effects are generated assuming a map of
event rates as a function of Galactic coordinates $(l,b)$. The probabilities
of lensing over the sky are drawn from the study of \citet{Han2008}. 
This probabilistic treatment is not based on the real existence of a modelled
lens close to the line of sight of the source, it simply uses the lensing
probability to randomly generate microlensing events for a given source during
the five year observing period.

The Einstein crossing time is also a function of the direction of
observation in the bulge, obtained from the same paper. The Einstein
time of the simulated events are drawn from a gaussian distribution
centered on the mean Einstein time. The impact parameter follows a
flat distribution from 0 to 1. The time of maximum is uniformely distributed
and completely random, from the beginning to the end of the mission.
The Paczynski formula \citep{Paczynski1986} is used to compute the light curve.

\subsubsection{Exoplanets}

One or two extra-solar planets are generated with distributions in
true mass $M_{p}$ and orbital period $P$ resembling those of \citet{Tabachnik2002}
which constitutes quite a reasonable approximation to the observed
distributions as of today, and extrapolated down to the masses close
to the mass of Earth. A detailed description can be found in \citet{Sozzetti2009}.

Semi-major axes are derived given the star mass, planet mass, and
period. Eccentricities are drawn from a power-law-type distribution,
where full circular orbits ($e=0.0$) are assumed for periods below
6 days. All orbital angles ($i$, $\omega$ and $\Omega$) are drawn
from uniform distributions. Observed correlations between different
parameters (e.g, $P$ and $M_{p}$ ) are reproduced.

Simple prescriptions for the radius (assuming a mass-radius relationship
from available structural models, see e.g. \citet{Baraffe2003}),
effective temperature, phase, and albedo (assuming toy models for
the atmospheres, such as a Lambert sphere) are provided, based on
the present-day observational evidence. 

For every dwarf star generated of spectral type between F and mid-K,
the likelihood that it harbours a planet of given mass and period
depends on its metal abundance according to the \citet{Fischer2005}
and \citet{Sozzetti2009} prescriptions. M dwarfs, giant stars, white
dwarfs, and young stars do not include simulated planets for the time
being, as well as double and multiple systems. 

The astrometric displacement, spectroscopic radial velocity amplitude,
and photometric dimming (when transiting) induced by a planet on the
parent star, and their evolution in time, are presently computed from
orbital components similarly to double stellar systems.

\subsection{Extragalactic objects}

\subsubsection{Resolved galaxies}

In order to simulate the Magellanic clouds, catalogues of stars and
their characteristics (magnitudes $B$, $V$, $I$, $T_\mathrm{eff}$, $log(g)$, spectral type)
have been obtained from the literature (Belcheva et al., private communication). 

For the astrometry, since star by star distance is missing, a single
proper motion and radial velocity for all stars of both clouds is
assumed. Chemical abundances are also guessed from the mean abundance
taken from the literature. The resulting values and their references
are given in table \ref{tab:Magellanic-clouds}. For simulating the
depth of the clouds a gaussian distribution is assumed along the line
of sight with a sigma given in the table. 

\begin{table*}
\begin{centering}
\caption{\label{tab:Magellanic-clouds}Assumed parameters of the Magellanic
clouds.}
\begin{tabular}{|c|c|c|c|l|}
\hline 
{\normalsize Parameter} & {\normalsize Units} & {\normalsize LMC} & {\normalsize SMC} & {\normalsize Reference}\tabularnewline
\hline 
\hline 
{\normalsize Distance} & {\normalsize kpc} & {\normalsize 48.1} & {\normalsize 60.6} & {\normalsize LMC: Macri et al. (2006) }\tabularnewline
 &  &  &  & {\normalsize SMC: Hilditch et al. (2005) }\tabularnewline
\hline 
{\normalsize Depth} & {\normalsize kpc} & {\normalsize 0.75} & {\normalsize 1.48} & {\normalsize LMC: Sakai et al. (2000) }\tabularnewline
 &  &  &  & {\normalsize SMC: Subramanian (2009) }\tabularnewline
\hline 
{\normalsize $\mu_{\alpha}\cos\left(\delta\right)$} & {\normalsize mas/yr} & {\normalsize 1.95} & {\normalsize 0.95} & {\normalsize Costa et al. (2009) }\tabularnewline
\hline 
{\normalsize $\mu_{\delta}$} & {\normalsize mas/yr} & {\normalsize 0.43} & {\normalsize -1.14} & {\normalsize Costa et al. (2009) }\tabularnewline
\hline 
{\normalsize $V_\text{los}$} & {\normalsize km/s} & {\normalsize 283} & {\normalsize 158} & {\normalsize SIMBAD (CDS) }\tabularnewline
\hline 
{\normalsize $[Fe/H$]} & {\normalsize dex} & {\normalsize -0.75$\pm$0.5} & {\normalsize -1.2$\pm$0.2} & {\normalsize Kontizas et al, (2011) }\tabularnewline
\hline 
{\normalsize $[\alpha/Fe]$} & {\normalsize dex} & {\normalsize 0.00$\pm$0.2} & {\normalsize 0.00$\pm$0.5} & {\normalsize Kontizas et al, (2011) }\tabularnewline
\hline 
\end{tabular}
\par\end{centering}

\end{table*}

Stellar masses are estimated for each star from polynomial fits of
the mass as a function of $B-V$ colour, for several ranges of $\log(g)$,
based on Padova isochrones for a metallicity of $z=0.003$ for the
LMC and $z=0.0013$ for the SMC. The gravities have been estimated
from the effective temperature and luminosity class but is very difficult
to assert from the available observables. Hence the resulting HR diagrams
for the Magellanic clouds are not as well defined and reliable as
they would be from theoretical isochrones.

\subsubsection{Unresolved galaxies}

Most galaxies observable by Gaia will not be resolved in their individual stars. These 
unresolved galaxies are simulated using the Stuff (catalog generation)
and Skymaker (shape/image simulation) codes from \cite{Bertin2009}, adapted
to Gaia by \cite{Dollet2004} and  \citep{Krone-Martins2008}.

This simulator generates a catalog of galaxies with a 2D uniform
distribution and a number density distribution in each Hubble type sampled from Schechter\textquoteright{}s
luminosity function \citep{Fioc1999}. Parameters of the luminosity
function for each Hubble type are given in table \ref{tab:Luminosity-functions-for-Galaxies}. 
Each galaxy is
assembled as a sum of a disc and a spheroid, they are located at their
redshift and luminosity and K corrections are applied. The algorithm
returns for each galaxy its position, magnitude, B/T relation, disc
size, bulge size, bulge flatness, redshift, position angles, and $V-I$. 

The adopted library of synthetic  spectra at low resolution has been created \cite{Tsalmantza2009} based on Pegase-2
code \citep{Fioc1997}(\url{http://www2.iap.fr/pegase}). 9 Hubble types are available (including Quenched
Star Forming Galaxies or QSFG) \cite{Tsalmantza2009}, with 5 different inclinations (0.00,
22.5, 45.0, 67.5, 90.0) for non-elliptical galaxies, and at 11 redshifts
(from 0. to 2. by step of 0.2). For all inclinations, Pegase-2 spectra
are computed with internal extinction by transfer model with two geometries
either slab or spheroid depending on type. 

It should be noted that the resulting percentages per type,  given 
in Table \ref{tab:Galaxies-by-type}, reflect the
numbers expected without applying the Gaia source detection and prioritization
algorithms. De facto, the detection efficiency will be better for unresolved galaxies 
having a prominent bulge and for nucleated galaxies, hence the effective Gaia catalog 
will have different percentages than the ones given in the table. 

\begin{table}
\begin{centering}
\caption{\label{tab:Luminosity-functions-for-Galaxies}Parameters defining
the luminosity function for different galaxy types at $z=0$, from
\citet{Fioc1999}. The LF follows a shape from \citet{Schechter1976}. 
M{*} (Bj)  is the magnitude in the Bj filter in the Schechter formalism.}
\begin{tabular}{|c|c|c|c|c|}
\hline 
Type & $\phi${*} ($Mpc^{-3}$) & M{*} (Bj) & Alpha & Bulge/Total\tabularnewline
\hline 
\hline 
E2 & $1.91\times10^{-3}$ & $-20.02$ & $-1$ & $1.0$\tabularnewline
\hline 
E-SO & $1.91\times10^{-3}$ & $-20.02$ & $-1$ & $0.9$\tabularnewline
\hline 
Sa & $2.18\times10^{-3}$ & $-19.62$ & $-1$ & $0.57$\tabularnewline
\hline 
Sb & $2.18\times10^{-3}$ & $-19.62$ & $-1$ & $0.32$\tabularnewline
\hline 
Sbc & $2.18\times10^{-3}$ & $-19.62$ & $-1$ & $0.32$\tabularnewline
\hline 
Sc & $4.82\times10^{-3}$ & $-18.86$ & $-1$ & $0.016$\tabularnewline
\hline 
Sd & $9.65\times10^{-3}$ & $-18.86$ & $-1$ & $0.049$\tabularnewline
\hline 
Im & $9.65\times10^{-3}$ & $-18.86$ & $-1$ & $0.0$\tabularnewline
\hline 
QSFG & $1.03\times10^{-2}$ & $-16.99$ & $-1.73$ & $0.0$\tabularnewline
\hline 
\end{tabular}
\par\end{centering}

\end{table}


\subsubsection{Quasars}

QSOs are simulated from the scheme proposed in \citet{Slezak2007}.
To summarize, lists of sources have been generated with similar statistical
properties as the SDSS, but extrapolated to G = 20.5 (the SDSS sample
being complete to i = 19.1) and taking into account the flatter slope
expected at the faint-end of the QSO luminosity distribution. The
space density per bin of magnitude and the luminosity function should
be very close to the actual sky distribution. Since bright quasars
are saturated in the SDSS, the catalogue is complemented by the \citet{Veroncetty2006}
catalogue of nearby QSOs. 

The equatorial coordinates have been generated from a uniform drawing
on the sphere in each of the sub-populations defined by its redshift.
No screening has been applied in the vicinity of the Galactic plane
since this will result directly from the application of the absorption
and reddening model at a later stage. Distance indicators can be derived
for each object from its redshift value by specifying a cosmological
model. Each of these sources lying at cosmological distances, a nearly
zero ($10^{-6}$mas) parallax has been assigned to all of them (equivalent
to an Euclidean distance of about 1 Gpc ) in order to avoid possible
overflow/underflow problems in the simulation.

In principle, distant sources are assumed to be co-moving with the
general expansion of the distant Universe and have no transverse motion.
However, the observer is not at rest with respect to the distant Universe
and the accelerated motion around the Galactic center, or more generally,
that of the Local group toward the Virgo cluster is the source of
a spurious proper motion with a systematic pattern. This has been
discussed in many places \citep{Kovalevsky2003,Mignard2005}. Eventually
the effect of the acceleration (centripetal acceleration of the solar
system) will show up as a small proper motion of the quasars, or stated
differently we will see the motion of the quasars on a tiny fraction
of the aberration ellipse whose period is 250 million years. This effect
is simulated directly in the quasar catalogue, and equations given
in \citet{Slezak2007}. This explains their not null proper motions
in the output catalogue.

\subsubsection{Supernovae}

A set of supernovae (SN) are generated associated with galaxies, with a proportion
for each Hubble type, as given in table \ref{tab:Probability-of-having-SN}. Numbers of SNIIs are
computed from the local star formation rate at 13Gyrs (z=0)
and IMF for M*(Bj) galaxy types as predicted by the code PEGASE.2.
Theoretical SNIa numbers follow the SNII/SNIa ratios from
\cite{Greggio1983}.
In this case, the SN is situated at a distance randomly selected at
less than a disc radius from the parent galaxy, accounting for the
inclination.

\begin{table}
\begin{centering}
\caption{\label{tab:Probability-of-having-SN}Predictions of SNII and SNIa numbers
exploding per century for M* (Bj) galaxies defined in table 2.11.}
\begin{tabular}{|c|c|c|}
\hline 
Hubble type & SNII /century & SNI /century\tabularnewline
\hline 
\hline 
E2 & 0.0 & 0.05 \tabularnewline
\hline 
E-S0 & 0.0 & 0.05 \tabularnewline
\hline 
Sa & 0.4 & 0.11 \tabularnewline
\hline 
Sb & 0.62 & 0.13 \tabularnewline
\hline 
Sbc & 1.07  & 0.16 \tabularnewline
\hline 
Sc & 0.16  & 0.07 \tabularnewline
\hline 
Sd & 0.049  & 0.07 \tabularnewline
\hline 
Im & 0.60  & 0.05 \tabularnewline
\hline 
QSFG & 0.0 & 0.05 \tabularnewline
\hline 
\end{tabular}
\par\end{centering}

\end{table}

Another set of SN are generated randomly on the sky to simulate SN
on host galaxies which are too faint in surface brightness to be detected
by Gaia.

Four SN types are available with a total probability of occuring
determined to give at the end 6366 SN per steradian during the 5 year
mission (from estimations by \citet{Belokurov2003}). For each SN generated
a type is attributed from the associated probability, and the absolute magnitude is computed 
from a Gaussian drawing centered on the absolute magnitude and the dispersion of the type corresponding to the cosmic scatter.
These parameters are given in table \ref{tab:Parameters-for-each-SN}.
Supernova light curves are from Peter Nugent%
\footnote{http://supernova.lbl.gov/\textasciitilde{}nugent/nugent\_templates.html%
}. It is assumed that the SN varies the same way at each wavelength
and the light curve in V has been taken as reference.

\begin{table}
\begin{centering}
\caption{\label{tab:Parameters-for-each-SN}Parameters for each supernova type taken 
from \citet{Belokurov2003}. The probability is given for each type, as well as the absolute magnitude 
and a dispersion about this magnitude corresponding to the cosmic variance.}
\begin{tabular}{|c|c|c|c|}
\hline 
Type & Probability & $M_{G}$ & Sigma\tabularnewline
\hline 
\hline 
Ia & 0.6663  & -18.99  & 0.76 \tabularnewline
\hline 
Ib/Ic & 0.0999  & -17.75  & 1.29 \tabularnewline
\hline 
II-L & 0.1978  & -17.63  & 0.88 \tabularnewline
\hline 
II-P & 0.0387  & -16.44  & 1.23 \tabularnewline
\hline 
\end{tabular}
\par\end{centering}

\end{table}

\subsection{Extinction model}

The extinction model, applied to Galactic and extragalatic objects,
is based on the dust distribution model of \citet{Drimmel2003}. This
full 3D extinction model is a strong improvement over previous generations
of extinction models as it includes both a smooth diffuse absorption
distribution for a disc and the spiral structure and smaller scale
corrections based on the integrated dust emission measured from the
far infrared (FIR). The extinction law is from \cite{Cardelli}.

\section{\label{sec:Gaia-Universe-Model-Snapshot}Gaia Universe Model Snapshot
(GUMS)}

The Gaia Universe Model Snapshot (GUMS) is part of the GOG component
of the Gaia simulator. It has been used to generate a synthetic catalogue
of objects from the universe model for a given static time $t_{0}$
simulating the real environment where Gaia will observe (down to $G=20$). 

It is worth noting that this snapshot is what Gaia will be able to
potentially observe but not what it will really detect, since satellite
instrument specifications and the available error models are not
 taken into account in the present statistical analysis. Gaia 
performances and error models are described at \url{http://www.rssd.esa.int/index.php?page=Science_Performance&project=GAIA}.

The generated universe model snapshot has been analyzed by using the
Gog Analysis Tool (GAT) statistics framework, which produces all types
of diagnostic statistics allowing its scientific validation (e.g.
star density distributions, HR diagrams, distributions of the properties
of the stars). The visual representation of the most interesting results
exposed in this article have been generated using Python, Healpy and
Matplotlib \citep{Hunter2007}.

The simulation was performed with MareNostrum, one of the most powerful
supercomputers in Europe managed by the Barcelona Supercomputing Center.
The execution took 20.000 hours (equivalent to 28 months) of computation
time distributed in 20 jobs, each one using between 16 and 128 CPUs.
MareNostrum runs a SUSE Linux Enterprise Server 10SP2 and its 2,560
nodes are powered by 2 dual-core IBM 64-bit PowerPC 970MP processors
running at 2.3 GHz.

\subsection{\label{sub:Overview}Galactic objects overview}
\begin{itemize}
\item G less than 20 mag
\end{itemize}
In general terms, the universe model has generated a total number
of 1,000,000,000 galactic objects of which \textasciitilde{}49\%
are single stars and \textasciitilde{}51\% stellar systems formed
by stars with planets and binary/multiple stars.

Individually, the model has created 1,600,000,000 stars where about 32\%
of them are single stars with magnitude G less than 20 (potentially
observable by Gaia) and 68\% correspond to stars in multiple
systems (see table \ref{tab:Overview-of-the-Universe-Model}). This
last group is formed by stars that have magnitude G less than 20
as a system but, in some cases, the isolated components can have magnitude
G superior to 20 and won't be individually detectable by Gaia. 

Only taking into consideration the magnitude limit in G and ignoring
the angular separation of multiple systems, Gaia could be able to
individually observe up to 1,100,000,000 stars (69\%)
in single and multiple systems.
\begin{itemize}
\item $G_\mathrm{RVS}$ less than 17 mag
\end{itemize}
For $G_\mathrm{RVS}$ magnitudes limited to 17, the model has generated 370,000,000
galactic objects of which \textasciitilde{}43\% are single stars
and \textasciitilde{}57\% stellar systems formed by stars with planets
and binary/multiple stars.

Concretely, the RVS instrument could potentially provide radial velocities
for up to 390,000,000 stars in single and multiple systems if the
limit in angular separation and resolution power are ignored (see table
\ref{tab:Overview-of-the-Universe-Model}).
\begin{itemize}
\item $G_\mathrm{RVS}$ less than 12 mag
\end{itemize}
The model has generated 13,100,000 galactic objects with
$G_\mathrm{RVS}$ less than 12, of which \textasciitilde{}27\% are single
stars and \textasciitilde{}73\% stellar systems formed by stars with
planets and binary/multiple stars.

Again, if the limits in angular separation and resolution power are
ignored, the RVS instrument could measure individual abundances of
key chemical elements (metallicities) for \textasciitilde{}13,000,000
stars (see table \ref{tab:Overview-of-the-Universe-Model}).

\begin{table}
\begin{centering}
\caption{\label{tab:Overview-of-the-Universe-Model}Overview of the number
of single stars and multiple system generated by the universe model.
Percentages have been calculated over the total stars.}
\begin{tabular}{|l|c|c|c|}
\hline 
{\scriptsize Stars} & {\scriptsize G < 20 mag} & {\scriptsize Grvs < 17 mag} & {\scriptsize Grvs < 12 mag}\tabularnewline
\hline 
\hline 
{\scriptsize Single stars} & {\scriptsize 31.59\%} & {\scriptsize 25.82\% } & {\scriptsize 12.91\% }\tabularnewline
\hline 
{\scriptsize Stars in multiple systems} & {\scriptsize 68.41\%} & {\scriptsize 74.18\% } & {\scriptsize 87.09\% }\tabularnewline
\hline 
\emph{\scriptsize $\Rightarrow$ In binary systems} & \emph{\scriptsize 52.25\%} & {\scriptsize 51.55\% } & \emph{\scriptsize 40.24\% }\tabularnewline
\hline 
\emph{\scriptsize $\Rightarrow$ Others (ternary, etc.)} & \emph{\scriptsize 16.16\%} & \emph{\scriptsize 22.63\% } & \emph{\scriptsize 46.85\% }\tabularnewline
\hline 
\hline 
{\scriptsize Total stars} & {\scriptsize 1,600,000,000} & {\scriptsize 600,000,000} & {\scriptsize 28,000,000 }\tabularnewline
\hline 
{\scriptsize Individually observable} & {\scriptsize 1,100,000,000} & {\scriptsize 390,000,000} & {\scriptsize 13,000,000}\tabularnewline
\hline 
\hline 
{\scriptsize $\Rightarrow$ }\emph{\scriptsize Variable} & {\scriptsize 1.78\% } & {\scriptsize 3.06\% } & {\scriptsize 8.37\% }\tabularnewline
\hline 
{\scriptsize $\Rightarrow$ }\emph{\scriptsize With planets} & {\scriptsize 1.75\% } & {\scriptsize 1.44\% } & {\scriptsize 0.66\% }\tabularnewline
\hline 
\end{tabular}
\par\end{centering}

\end{table}

\subsection{Star distribution}

The distribution of stars in the sky has been plotted using a Hierarchical
Equal Area isoLatitude Pixelisation, also known as Healpix projection.
Unlike the HTM internal representation of the sky explained in section
\ref{sec:Gaia-simulator}, Healpix provide areas of identical size
which are useful for comparison.

\begin{figure}[ht!]
\begin{centering}
 \includegraphics[scale=0.55]{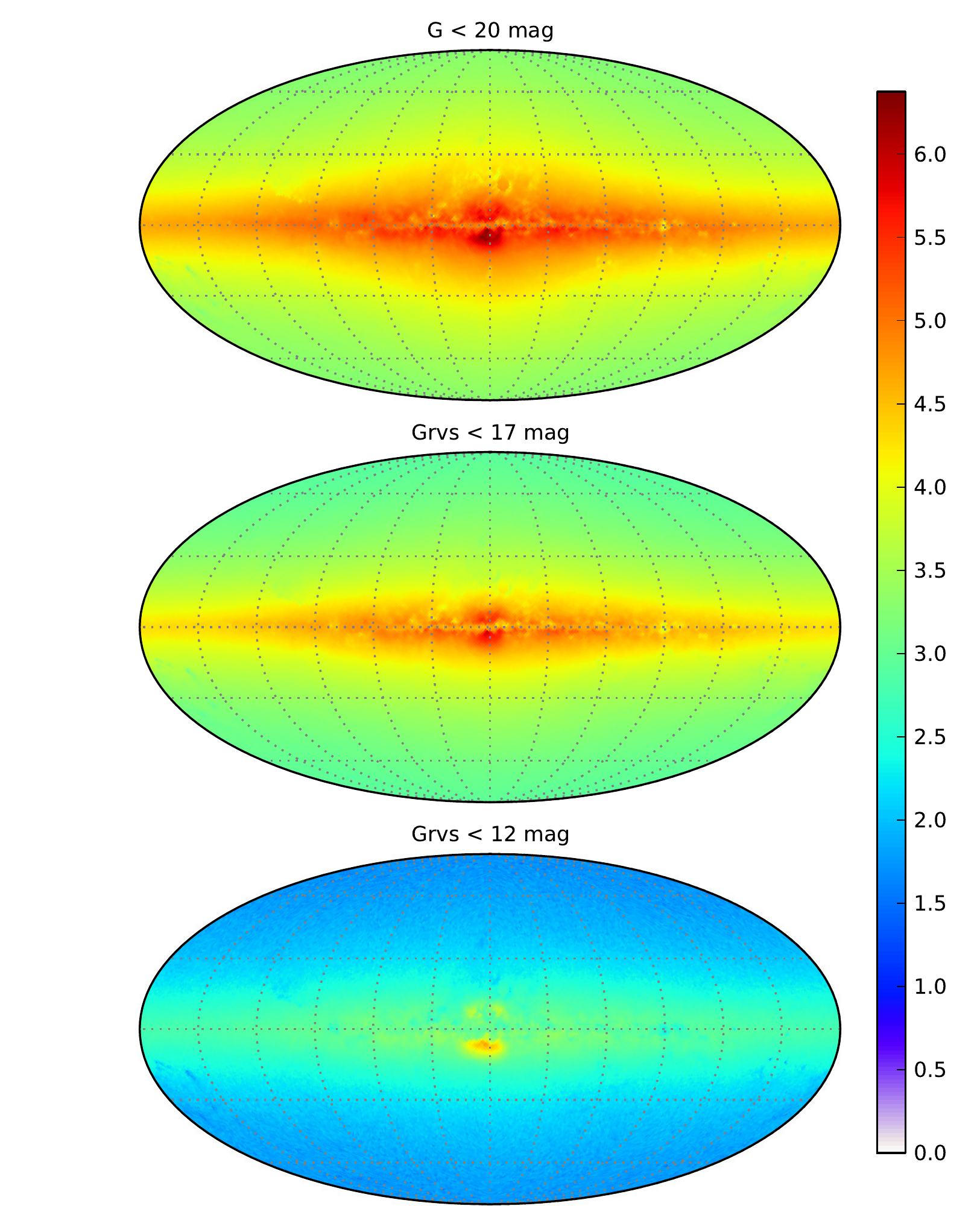}
\par\end{centering}

\caption{\label{fig:Total-sky-distribution-stars}Total sky distribution of
stars for different magnitudes. Top down: $G<20$, $G_\mathrm{RVS}<17$ and
$G_\mathrm{RVS}<12$. Color scale indicates the $\log_{10}$ of the number
of stars per square degree.}
\end{figure}

\begin{figure*}
\begin{centering}
\includegraphics[scale=1]{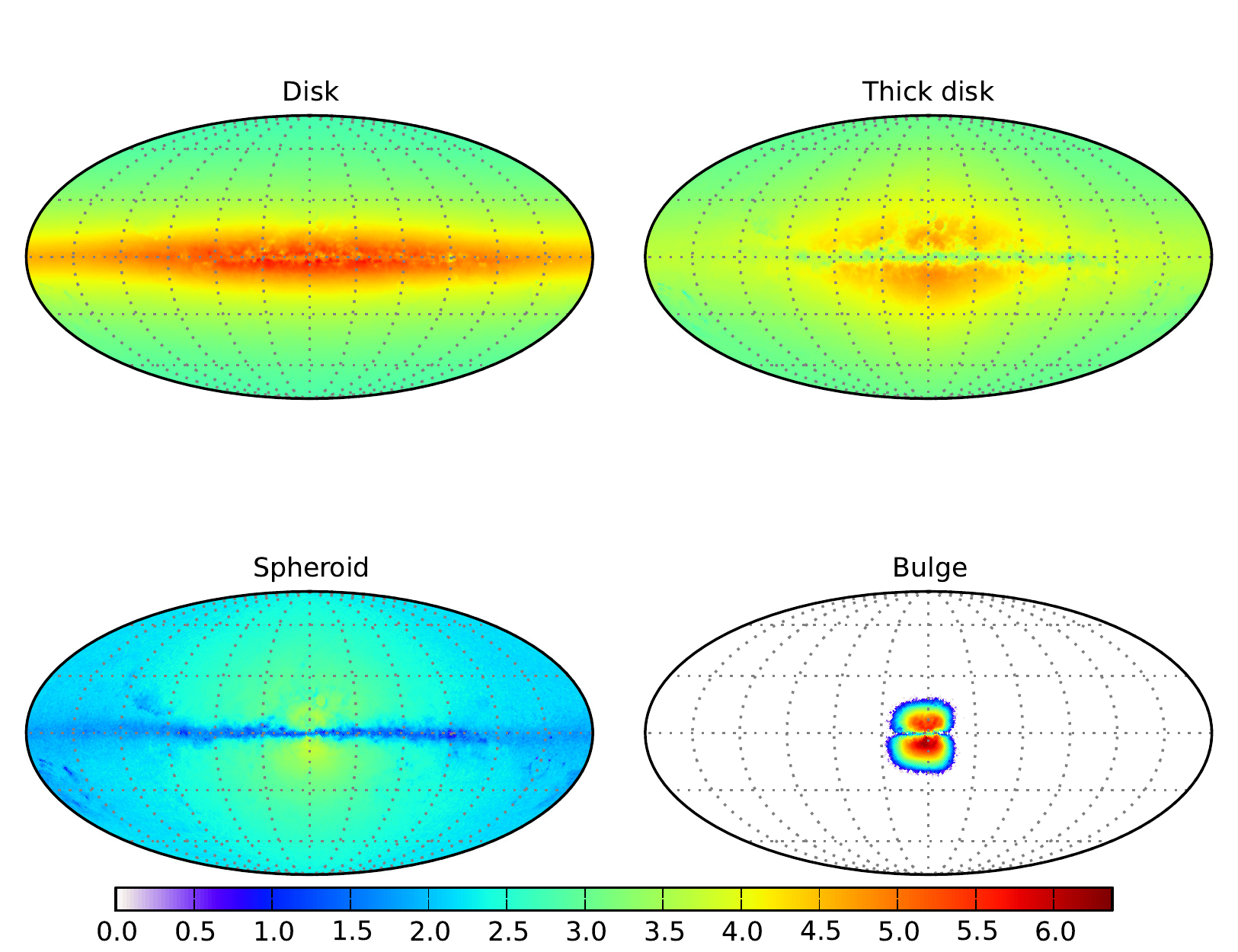}
\par\end{centering}

\caption{\label{fig:Stellar-distribution-by-pop}Stellar distribution split
by population ($G<20$). Left to right, top down: thin disc, thick
disc, spheroid and bulge. Color scale indicates the $\log_{10}$ of
the number of stars per square degree.}
\end{figure*}

The number of stars on every region of the sky varies significantly
depending on the band (figure \ref{fig:Total-sky-distribution-stars})
and the population (figure \ref{fig:Stellar-distribution-by-pop}).
In this last case, it is clear how the galactic center is concentrated
in the middle of the galaxy with only 10\% of stars, while the thin
disc is the densest region (67\%) as stated in table \ref{tab:Populations}.

The effects of the extinction model due to the interstellar material
of the Galaxy, predominantly atomic and molecular hydrogen and significant
amounts of dust, is clearly visible in these representations of the
sky.

\begin{table}[ht!]
\begin{centering}
\caption{\label{tab:Populations}Stars by population. Percentages have been
calculated over the total number of stars for each respective column.}
\begin{tabular}{|c|c|c|c|}
\hline 
{\scriptsize Population} & {\scriptsize G < 20 mag} & {\scriptsize Grvs < 17 mag} & {\scriptsize Grvs < 12 mag}\tabularnewline
\hline 
\hline 
{\scriptsize disc} & {\scriptsize 66.59\% } & {\scriptsize 76.82\% } & {\scriptsize 76.21\% }\tabularnewline
\hline 
{\scriptsize Thick disc} & {\scriptsize 21.88\% } & {\scriptsize 14.39\% } & {\scriptsize 8.75\% }\tabularnewline
\hline 
{\scriptsize Spheroid} & {\scriptsize 1.25\% } & {\scriptsize 0.58\% } & {\scriptsize 0.19\% }\tabularnewline
\hline 
{\scriptsize Bulge} & {\scriptsize 10.28\% } & {\scriptsize 8.22\% } & {\scriptsize 14.85\% }\tabularnewline
\hline 
\hline 
{\scriptsize Total} & {\scriptsize 1,100,000,000 } & {\scriptsize 390,000,000 } & {\scriptsize 13,000,000 }\tabularnewline
\hline 
\end{tabular}
\par\end{centering}

\end{table}

\begin{figure*}
\begin{centering}
\includegraphics[scale=0.7]{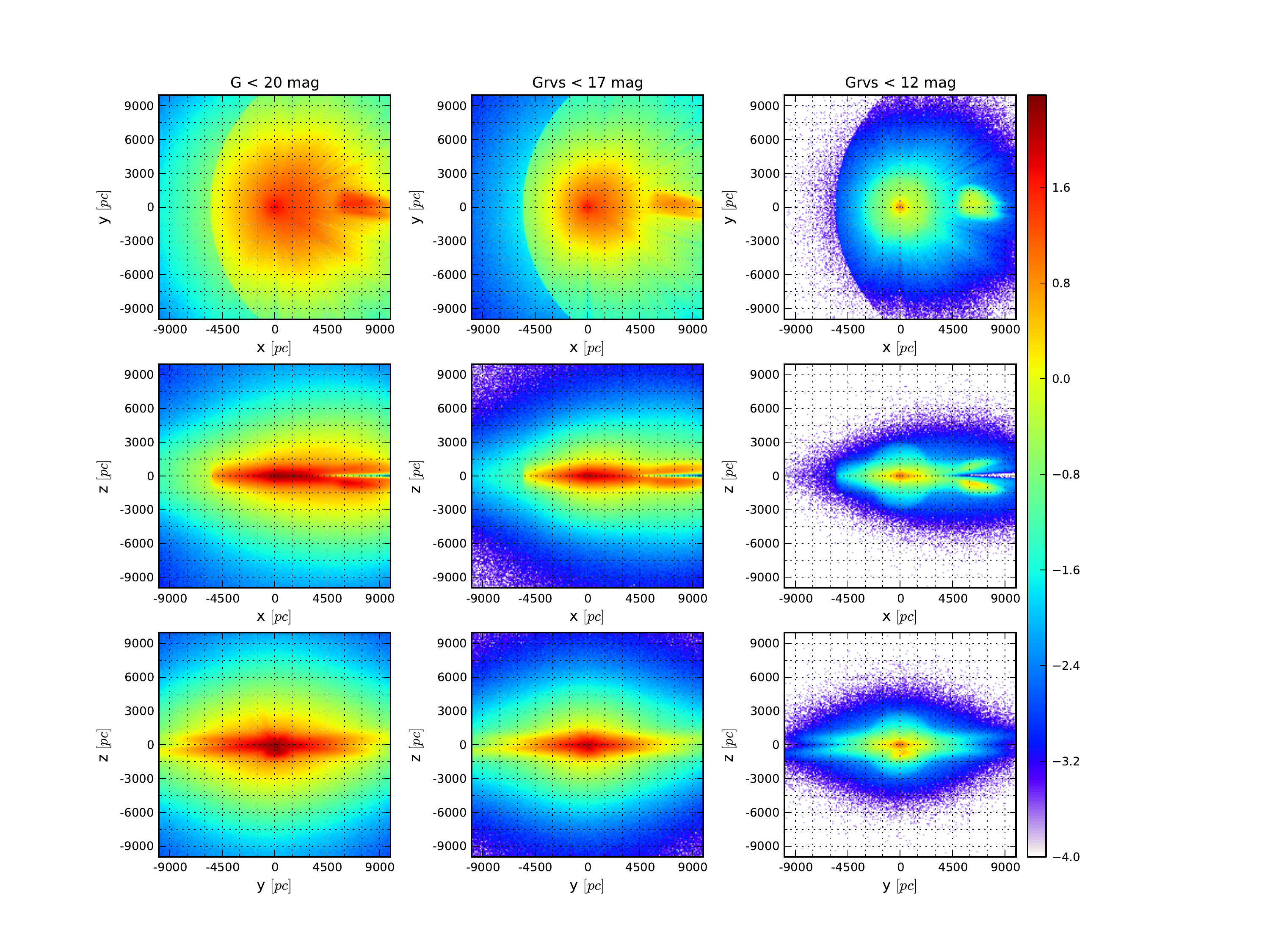}
\par\end{centering}

\caption{\label{fig:Distribution-of-star-XYZ}Distribution of star in heliocentric
Cartesian coordinates ($G<20$): top (XY), side (XZ) and front (YZ)
perspectives. Color scale indicates the $\log_{10}$ of the number
of stars per square parsec.}
\end{figure*}

A projected representation in heliocentric-galactic coordinates of
the stellar distribution shows how the majority of the generated stars
are denser near the sun position, located at the origin of the XYZ
coordinate system, and the bulge at 8.5 kpc away (figure \ref{fig:Distribution-of-star-XYZ}).
Specially from a top perspective (XY view), it is appreciable the
extinction effect which produces windows where farther stars can be
observed. One also notice the sudden density drop towards the anticenter which 
is due to the edge of the disc, assumed to be at a galactocentric distance of 14 kpc, following \citet{Robin1992}.

\begin{figure}[hb!]
\begin{centering}
\includegraphics[scale=0.5]{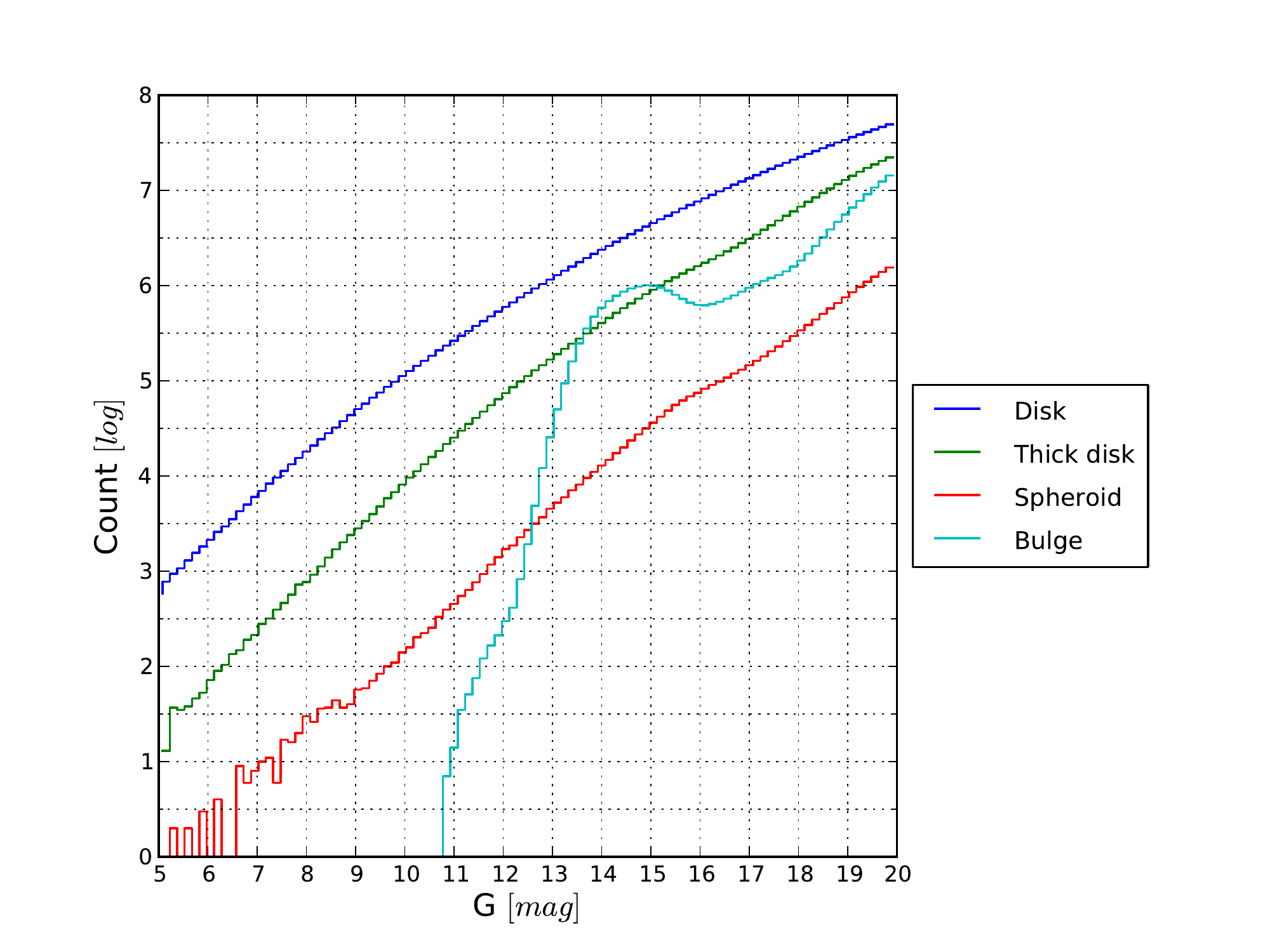}
\par\end{centering}

\caption{\label{fig:G-distribution}$G$ distribution split by stellar population.
It is worth noting that the bump for the bulge is due to the red clump,
which is seen at I=15 in Baade's window. }

\end{figure}

The distribution of stars according to the G magnitude varies depending
on the stellar population, being specially different for the bulge
(figure \ref{fig:G-distribution}).

\subsection{Star classification}

As expected, the most abundant group of stars belong to the main sequence
class (69\%), followed by sub-giants (15\%) and giants (14\%).
The complete star luminosity classification is given in table \ref{tab:Luminosity-class-of-stars}.

\begin{figure}
\begin{centering}
\includegraphics[scale=0.45]{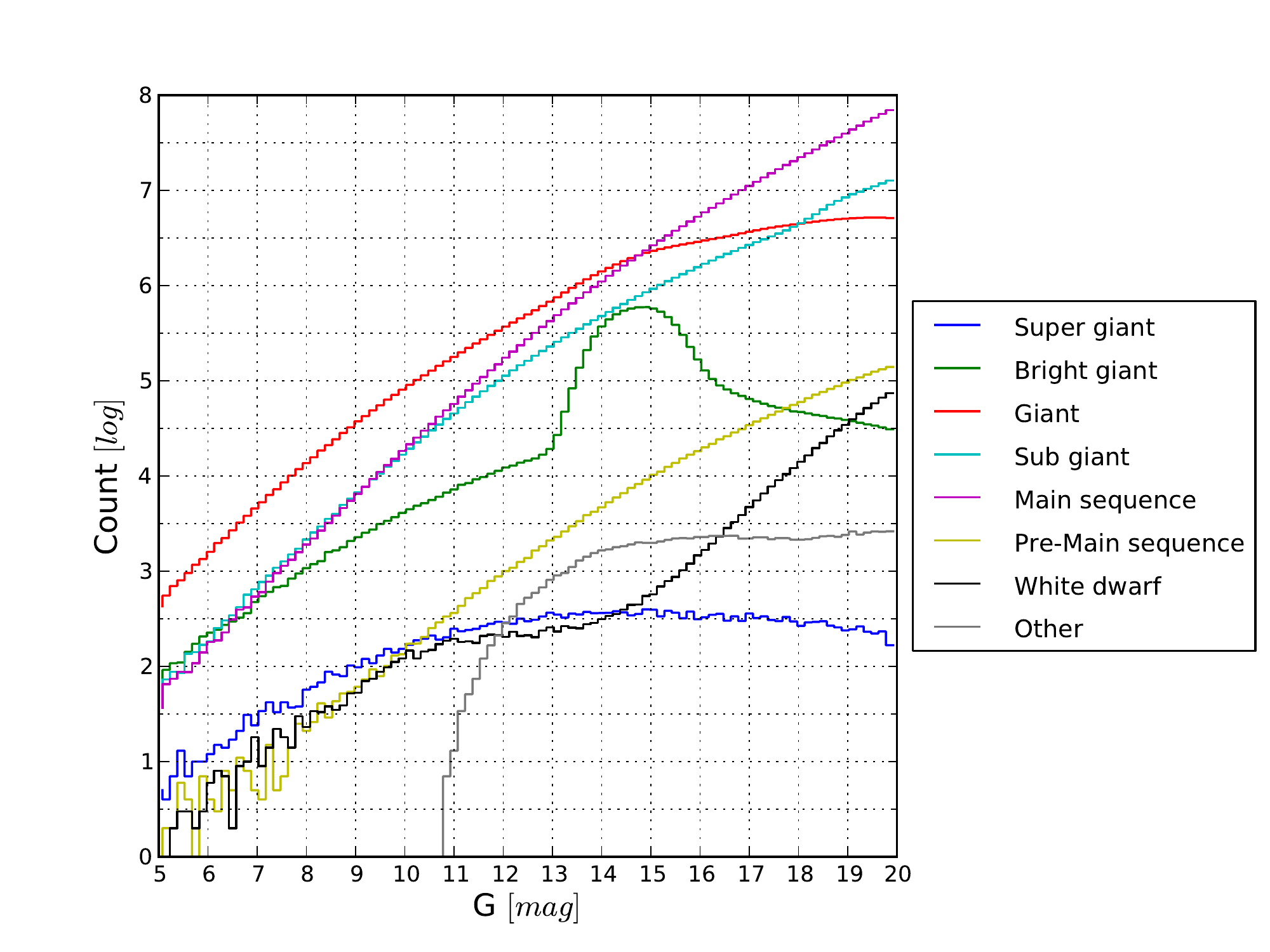}
\par\end{centering}

\caption{\label{fig:Star-distribution-split-by-lumClass}Star distribution
split by luminosity class for $G<20$. }
\end{figure}

The star distribution as a function of $G$ can be found in figure \ref{fig:Star-distribution-split-by-lumClass}.
Main sequence stars present the biggest exponential increase, relatively
similar to sub giants. The population of white dwarfs increases significantly
starting at magnitude $G=14$. It is also interesting how supergiants
decrease in number because they are intrinsically so bright and the peak that bright
giants present at $G=14.5$. For both of them, the decrease corresponds mainly to the distance of the edge of the disc in the Galactic plane.

\begin{table}
\begin{centering}
\caption{\label{tab:Luminosity-class-of-stars}Luminosity class of generated
stars. Percentages have been calculated over the total number of stars
for each respective column.}
\begin{tabular}{|c|c|c|c|}
\hline 
{\scriptsize Luminosity class} & {\scriptsize G < 20 mag} & {\scriptsize Grvs < 17 mag} & {\scriptsize Grvs < 12 mag}\tabularnewline
\hline 
\hline 
{\scriptsize supergiant} & {\scriptsize 0.00\% } & {\scriptsize 0.01\% } & {\scriptsize 0.07\% }\tabularnewline
\hline 
{\scriptsize Bright giant} & {\scriptsize 0.81\% } & {\scriptsize 2.18\% } & {\scriptsize 11.01\% }\tabularnewline
\hline 
{\scriptsize Giant} & {\scriptsize 14.47\% } & {\scriptsize 28.38\% } & {\scriptsize 62.71\% }\tabularnewline
\hline 
{\scriptsize Sub-giant} & {\scriptsize 15.08\% } & {\scriptsize 14.38\% } & {\scriptsize 10.32\% }\tabularnewline
\hline 
{\scriptsize Main sequence} & {\scriptsize 69.40\% } & {\scriptsize 54.82\% } & {\scriptsize 15.76\% }\tabularnewline
\hline 
{\scriptsize Pre-main sequence} & {\scriptsize 0.18\% } & {\scriptsize 0.20\% } & {\scriptsize 0.08\% }\tabularnewline
\hline 
{\scriptsize White dwarf} & {\scriptsize 0.05\% } & {\scriptsize 0.01\% } & {\scriptsize 0.03\% }\tabularnewline
\hline 
{\scriptsize Others} & {\scriptsize 0.01\% } & {\scriptsize 0.02\% } & {\scriptsize 0.02\% }\tabularnewline
\hline 
\hline 
{\scriptsize Total} & {\scriptsize 1,100,000,000} & {\scriptsize 390,000,000 } & {\scriptsize 13,000,000 }\tabularnewline
\hline 
\end{tabular}
\par\end{centering}

\end{table}

The spectral classification of stars (table \ref{tab:Spectral-types})
shows that G types are the most numerous (38\%), followed by K
types (28\%) and F types (23\%).

\begin{table}
\begin{centering}
\caption{\label{tab:Spectral-types}Spectral types of generated stars. Percentages
have been calculated over the total number of stars for each respective
column.}
\begin{tabular}{|c|c|c|c|}
\hline 
{\scriptsize Spectral type} & {\scriptsize G < 20 mag} & {\scriptsize Grvs < 17 mag} & {\scriptsize Grvs < 12 mag}\tabularnewline
\hline 
\hline 
{\scriptsize O} & {\scriptsize <0.01\%} & {\scriptsize <0.01\%} & {\scriptsize <0.01\% }\tabularnewline
\hline 
{\scriptsize B} & {\scriptsize 0.26\%} & {\scriptsize 0.50\% } & {\scriptsize 0.88\% }\tabularnewline
\hline 
{\scriptsize A} & {\scriptsize 1.85\%} & {\scriptsize 3.30\% } & {\scriptsize 4.84\% }\tabularnewline
\hline 
{\scriptsize F} & {\scriptsize 23.13\%} & {\scriptsize 22.94\% } & {\scriptsize 13.83\% }\tabularnewline
\hline 
{\scriptsize G} & {\scriptsize 38.28\%} & {\scriptsize 31.58\% } & {\scriptsize 15.46\% }\tabularnewline
\hline 
{\scriptsize K} & {\scriptsize 27.68\%} & {\scriptsize 32.23\% } & {\scriptsize 41.75\% }\tabularnewline
\hline 
{\scriptsize M} & {\scriptsize 7.75\%} & {\scriptsize 6.78\% } & {\scriptsize 11.38\% }\tabularnewline
\hline 
{\scriptsize L} & {\scriptsize <0.01\%} & {\scriptsize <0.01\% } & {\scriptsize <0.01\% }\tabularnewline
\hline 
{\scriptsize WR} & {\scriptsize <0.01\%} & {\scriptsize <0.01\% } & {\scriptsize 0.01\% }\tabularnewline
\hline 
{\scriptsize AGB} & {\scriptsize 0.91\%} & {\scriptsize 2.50\% } & {\scriptsize 11.37\% }\tabularnewline
\hline 
{\scriptsize Other} & {\scriptsize 0.09\%} & {\scriptsize 0.07\% } & {\scriptsize 0.33\% }\tabularnewline
\hline 
\hline 
{\scriptsize Total} & {\scriptsize 1,100,000,000} & {\scriptsize 390,000,000 } & {\scriptsize 13,000,000 }\tabularnewline
\hline 
\end{tabular}
\par\end{centering}

\end{table}

\begin{figure*}
\begin{centering}
\includegraphics[scale=0.75]{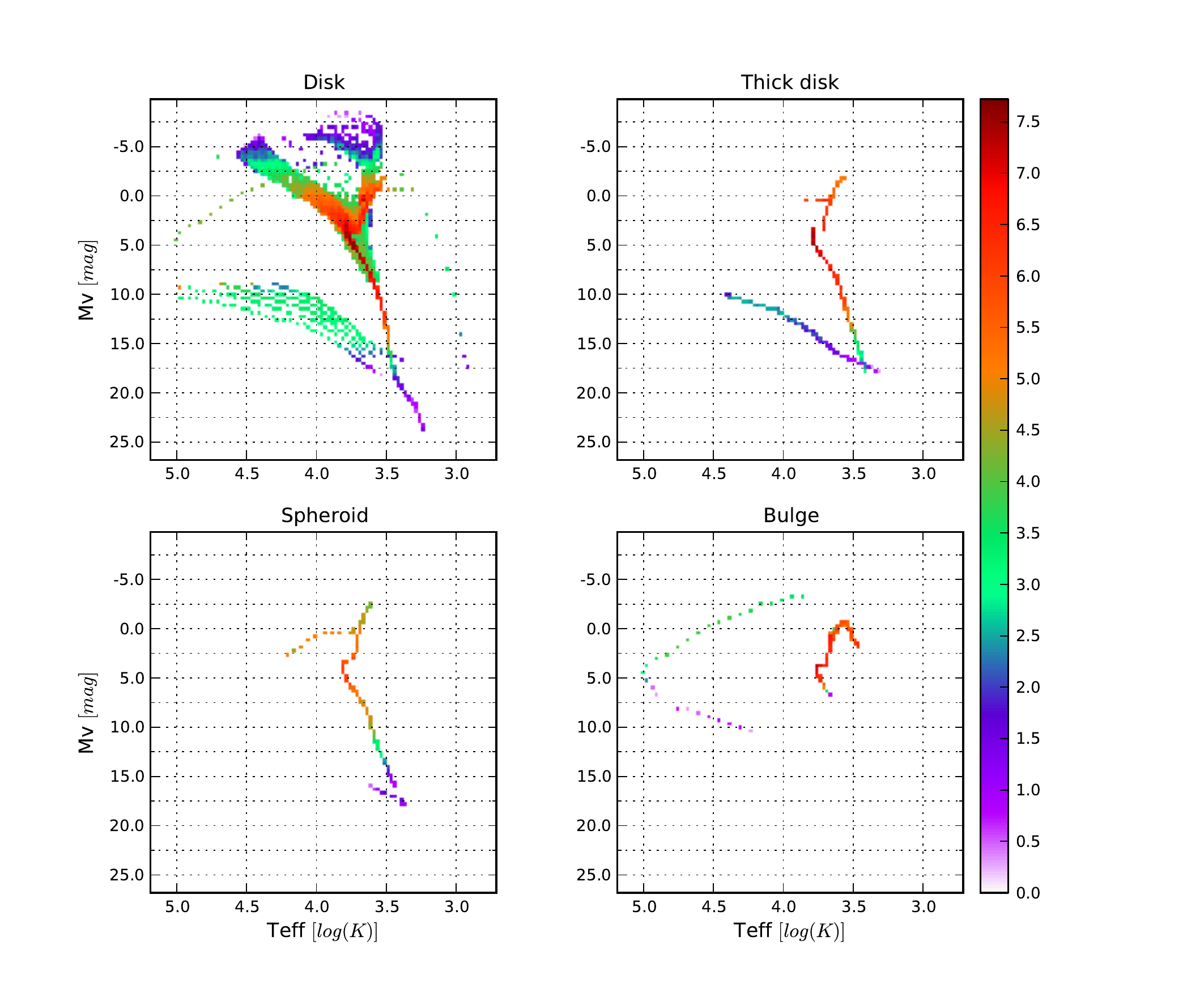}
\par\end{centering}

\caption{\label{fig:HR-Diagram-pop}HR Diagram of stars split by population
(left to right, top down): thin disc, thick disc, spheroid and bulge.
Color scale indicates the $\log_{10}$ of the number of stars per
0.025 $\log\left(K\right)$ and 0.37 mag.}

\end{figure*}

Considering the temperature and magnitude relation, HR diagrams have
been generated (figure \ref{fig:HR-Diagram-pop}). Disc population
represents the most complete one in terms of luminosity classes and
spectral types distribution. The thin disc sequence of AGB 
stars going roughly from -4 to +17 in Mv is due to the heavy internal 
reddening of this population in visual bands. Isochrones can be clearly identified
for thick disc, spheroid and bulge populations because they are assumed 
single burst generation. Additionally, all populations present a fraction of the white dwarfs generated by the model. 

\begin{figure}
\begin{centering}
\includegraphics[scale=0.4]{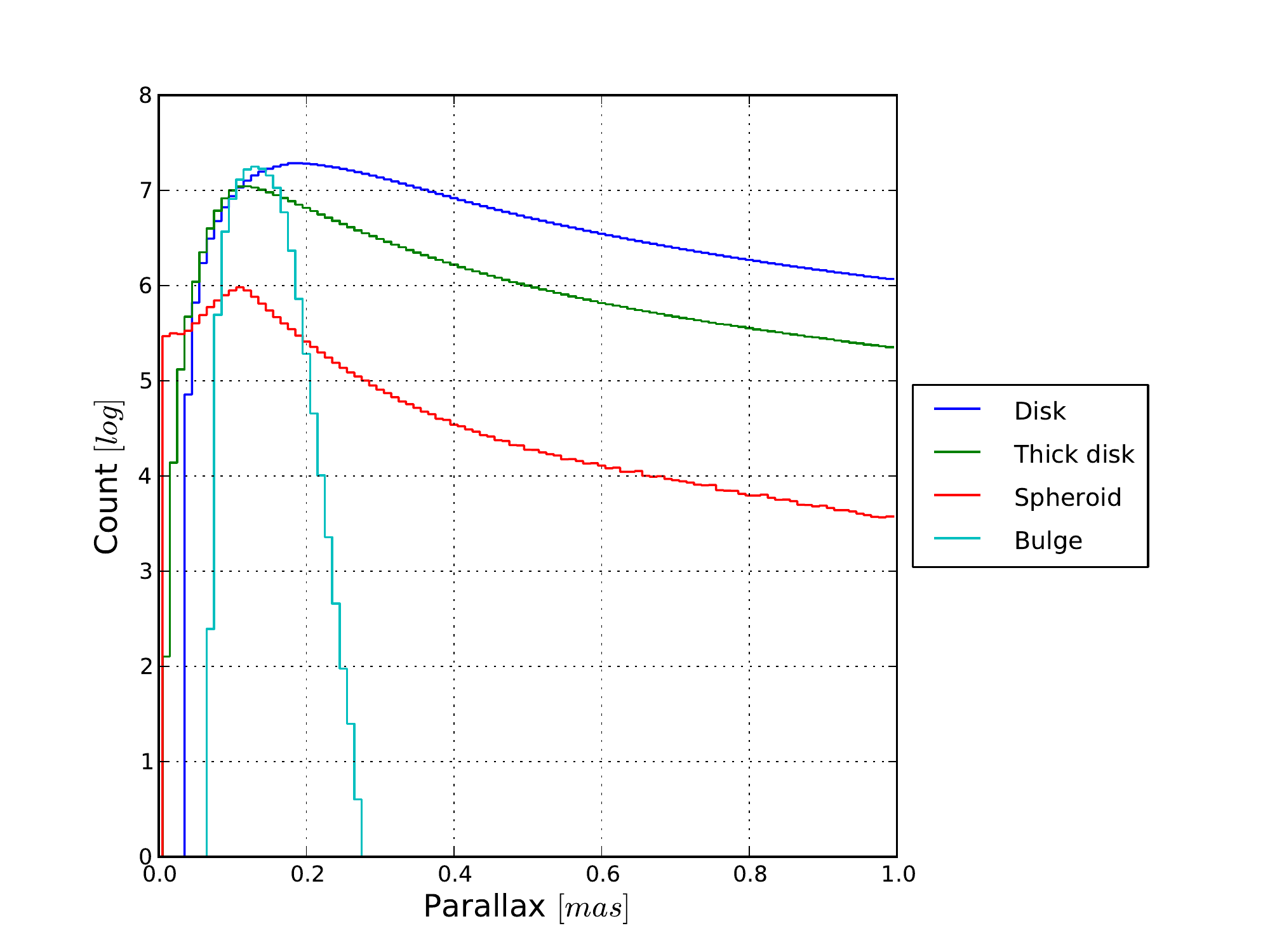}
\par\end{centering}

\caption{\label{fig:Stellar-parallax-pop}Stellar parallax split by populations}

\end{figure}

From the population perspective (figure \ref{fig:Stellar-parallax-pop}),
its is natural that bulge stars are concentrated at smaller parallaxes
(bigger distances) while the rest increases from the sun position
until the magnitude limit is reached and only brighter stars can be
observed. 

\begin{figure}
\begin{centering}
\includegraphics[scale=0.4]{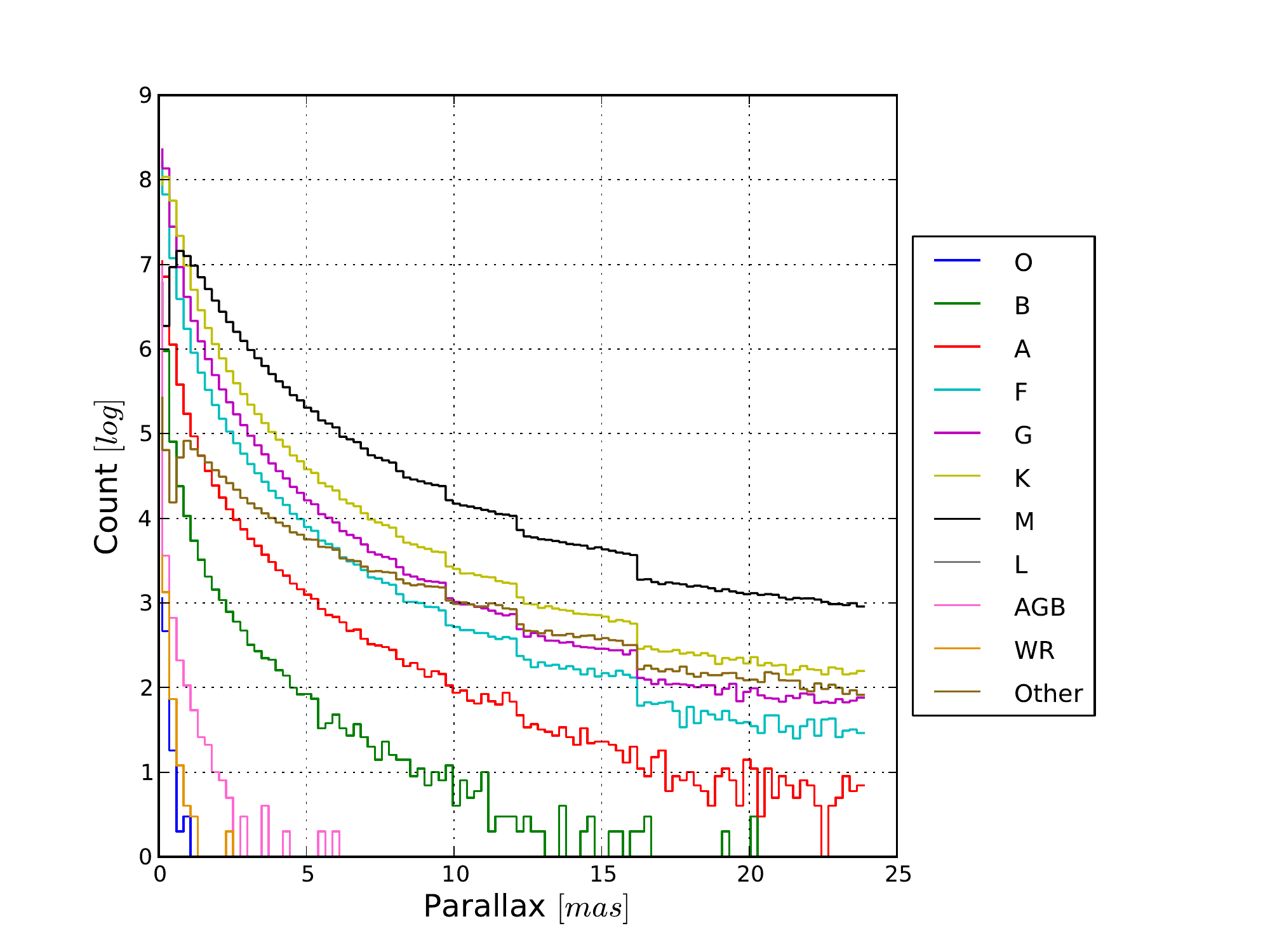}
\par\end{centering}

\caption{\label{fig:Stellar-parallax-spectral}Stellar parallax split by spectral
type}
\end{figure}

On the other hand, from a spectral type perspective, figure \ref{fig:Stellar-parallax-spectral}
presents the different parallax distributions for each type. Concrete
numbers can be found in table \ref{tab:Spectral-type-Parallaxes}
for specific parallaxes 240 $\mu$as, 480 $\mu$as and 960 $\mu$as%
\footnote{The choice of these somewhat odd values of parallax comes from limitation
of the available GAT statistics.%
} which correspond to distances 4167 pc, 2083 pc and 1042 pc.

\begin{table}
\begin{centering}
\caption{\label{tab:Spectral-type-Parallaxes}Number of stars for each spectral
type at different parallaxes. Percentages have been calculated over
totals per spectral type, which can be deduced from table \ref{tab:Spectral-types}.}
\begin{tabular}{|c|c|c|c|}
\hline 
{\scriptsize Spectral type} & {\scriptsize $\pi>240\mu$as} & {\scriptsize $\pi>480\mu$as} & {\scriptsize $\pi>960\mu$as}\tabularnewline
\hline 
\multirow{1}{*}{{\scriptsize O}} & {\scriptsize 30.25\%} & {\scriptsize 1.54\%} & {\scriptsize 0.31\%}\tabularnewline
\cline{2-4} 
\multirow{1}{*}{{\scriptsize B}} & {\scriptsize 38.38\%} & {\scriptsize 4.69\%} & {\scriptsize 0.99\%}\tabularnewline
\cline{2-4} 
\multirow{1}{*}{{\scriptsize A}} & {\scriptsize 45.61\%} & {\scriptsize 9.87\%} & {\scriptsize 2.33\%}\tabularnewline
\cline{2-4} 
\multirow{1}{*}{{\scriptsize F}} & {\scriptsize 35.06\%} & {\scriptsize 8.07\%} & {\scriptsize 1.74\%}\tabularnewline
\cline{2-4} 
\multirow{1}{*}{{\scriptsize G}} & {\scriptsize 44.52\%} & {\scriptsize 11.48\% } & {\scriptsize 2.46\%}\tabularnewline
\cline{2-4} 
\multirow{1}{*}{{\scriptsize K}} & {\scriptsize 70.61\%} & {\scriptsize 34.28\%} & {\scriptsize 7.94\%}\tabularnewline
\cline{2-4} 
\multirow{1}{*}{{\scriptsize M}} & {\scriptsize 92.75\%} & {\scriptsize 90.50\%} & {\scriptsize 62.09\% }\tabularnewline
\cline{2-4} 
\multirow{1}{*}{{\scriptsize L}} & {\scriptsize 0.00\%} & {\scriptsize 0.00\%} & {\scriptsize 0.00\%}\tabularnewline
\cline{2-4} 
\multirow{1}{*}{{\scriptsize WR}} & {\scriptsize 27.98\%} & {\scriptsize 1.89\%} & {\scriptsize 0.23\%}\tabularnewline
\cline{2-4} 
\multirow{1}{*}{{\scriptsize AGB}} & {\scriptsize 0.05\%} & {\scriptsize 0.01\%} & {\scriptsize <0.01\%}\tabularnewline
\cline{2-4} 
\multirow{1}{*}{{\scriptsize Other}} & {\scriptsize 71.56\%} & {\scriptsize 64.73\%} & {\scriptsize 57.53\%}\tabularnewline
\cline{2-4} 
\multirow{1}{*}{{\scriptsize Total}} & {\scriptsize 570,000,000 } & {\scriptsize 250,000,000 } & {\scriptsize 90,000,000 }\tabularnewline
\hline 
\end{tabular}
\par\end{centering}

\end{table}

\begin{figure*}
\begin{centering}
\includegraphics[scale=0.7]{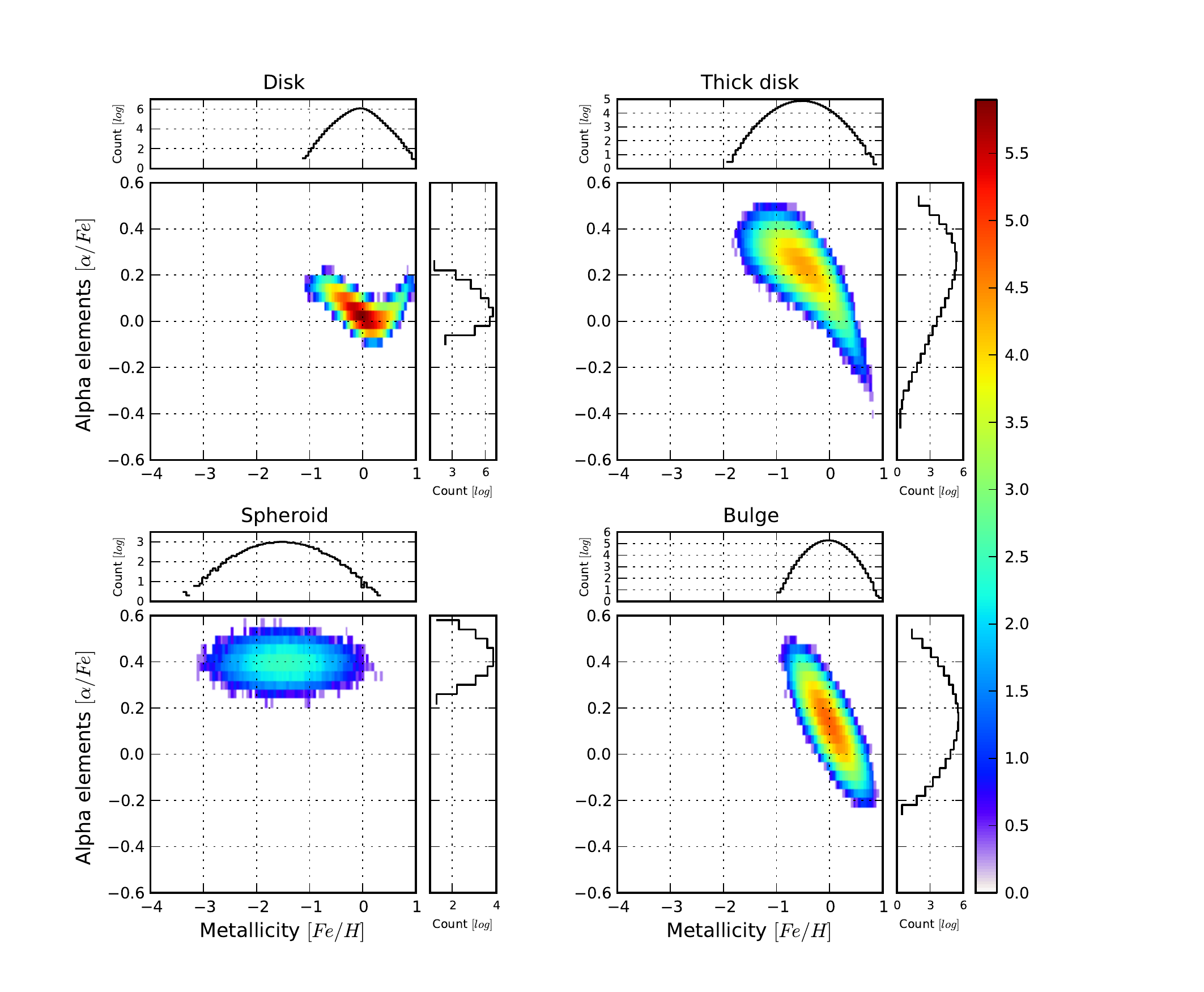}
\par\end{centering}

\caption{\label{fig:Metallicity-Alpha-elements}Metallicity - alpha elements
relation for $G_\mathrm{RVS}<12$. Color scale indicates the $\log_{10}$
of the number of stars per 0.05 $\left[Fe/H\right]$ and 0.04 $\left[\alpha/Fe\right]$.}
\end{figure*}

As mentioned in the introduction, the RVS instrument will be able
to measure abundances of key chemical elements (e.g. Ca, Mg and Si)
for stars up to $G_\mathrm{RVS}=12$. Metallicities and its relation to the
abundances of alpha elements is presented in figure \ref{fig:Metallicity-Alpha-elements}
split by population. Alpha element abundances are mostly reliable except at
metallicity larger than 0.5, due to the formulation which is extrapolated. It will be corrected in a future version.

\subsection{Kinematics}

\begin{figure*}
\begin{centering}
\includegraphics[scale=0.7]{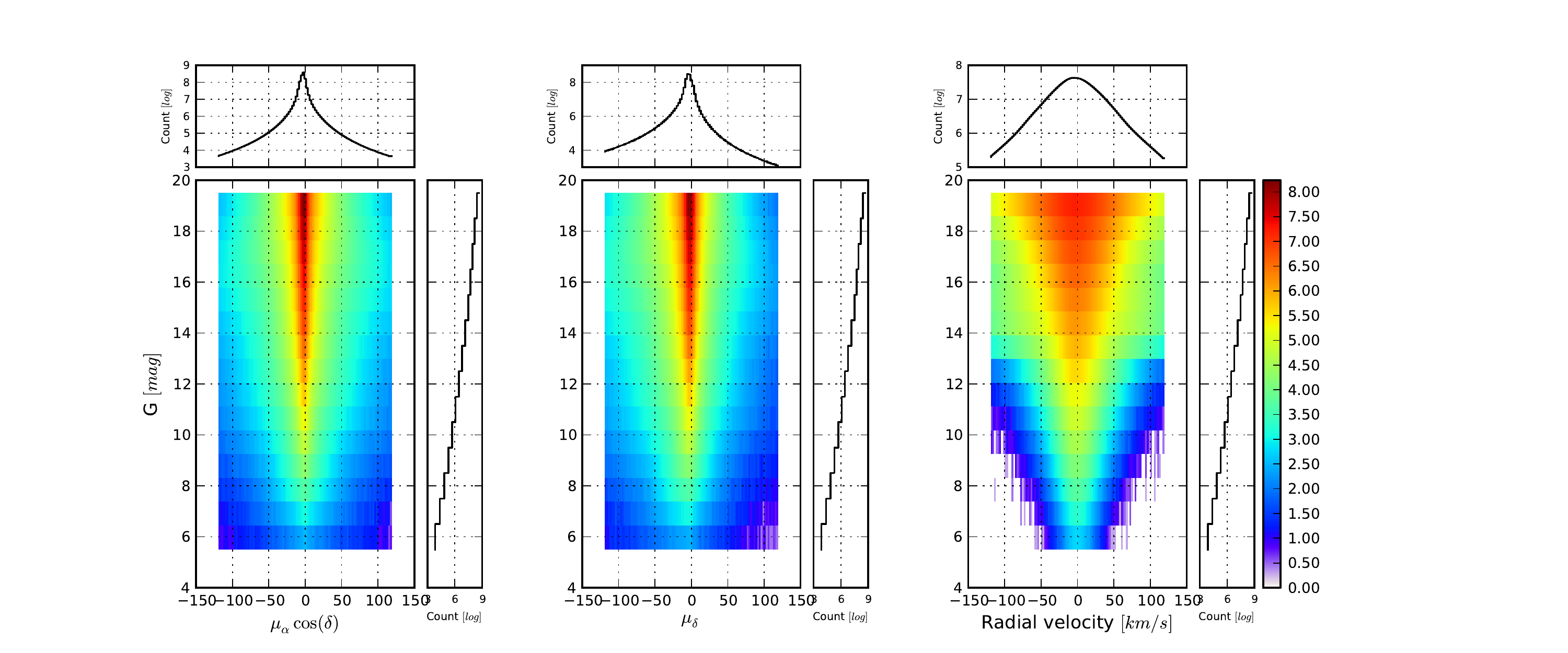}
\par\end{centering}

\caption{\label{fig:Proper-motion-Radial-vel}Proper motion of stars $\mu_{\alpha}cos\delta$,
$\mu_{\delta}$ and radial velocity $V_{R}$. Color scale indicates
the $\log_{10}$ of the number of stars per 2.4 mas/year (km/s in
case of $V_{R}$) and 1.0 mag.}

\end{figure*}

Proper motion of stars and radial velocity are represented in figure
\ref{fig:Proper-motion-Radial-vel}. Means are located at $\mu_{\alpha}\cos\left(\delta\right)=-1.95$ mas/year
and $\mu_{\delta}=-2.78$ mas/year, which are affected by the motion
of the solar local standard of rest.

\begin{figure*}
\begin{centering}
\includegraphics[scale=0.8]{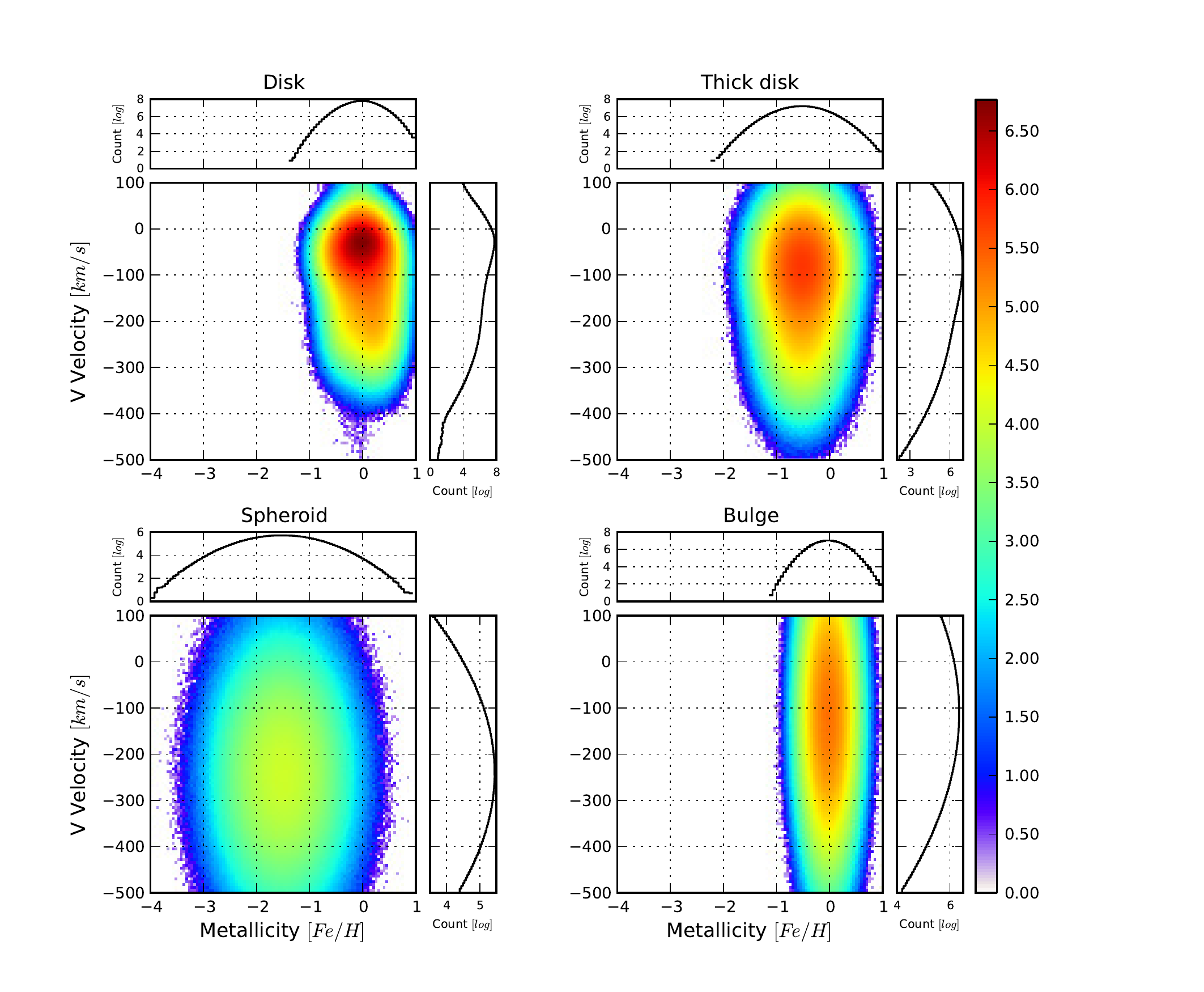}
\par\end{centering}

\caption{\label{fig:Metalicity-and-V-axis-vel}Metallicity and V axis velocity
relation split by population (left to right, top down): thin disc,
thick disc, spheroid and bulge. Color scale indicates the $\log_{10}$
of the number of stars per 0.05 $\left[Fe/H\right]$ and 6 km/s. }

\end{figure*}

Older stars, which present poorest metallicities,
tend to have lowest velocities on the V axis compared to the solar
local standard of rest (figure \ref{fig:Metalicity-and-V-axis-vel}) following the so-called asymmetric drift.
The approximate mean V velocities are -48 km~s$^{-1}$ for the
thin disc, -98 km~s$^{-1}$ for the thick disc, -243 km~s$^{-1}$
for the spheroid and -116 km~s$^{-1}$ for the bulge.

\subsection{Variable stars}

From the total amount of 1,600,000,000 individual stars generated
by the model, \textasciitilde{}1.8\% are variable stars (of the variable types included in the Universe Model).
They are composed by 6,800,000 single stars (\textasciitilde{}25\%
over total variable stars) with magnitude G less than 20 (observable
by Gaia if their maximum magnitude is reached at least once during the mission) 
and 21,000,000 stars in multiple system (\textasciitilde{}75\%).
However, as explained in section \ref{sub:Overview}, this last group
is formed by stars that have magnitude G less than 20 as a system
but, in some cases, its isolated components can have magnitude G superior
to 20 and won't be individually detectable by Gaia.

Again, only taking into consideration magnitude G and ignoring the
angular separation of multiple systems, Gaia could be able to observe
up to 21,500,000 variable stars in single and multiple systems (\textasciitilde{}2\%
over total individually observable stars exposed in section \ref{sub:Overview}).

Regarding radial velocities, they will be measurable for 16,000,000
variable stars with $G_\mathrm{RVS}<17$, while metallicities
will be available for 2,000,000 variable stars (\textasciitilde{}1\%)
with $G_\mathrm{RVS}<12$ (see table \ref{tab:Overview-of-variable}).

\begin{table}
\begin{centering}
\caption{\label{tab:Overview-of-variable}Overview of variable stars. Percentages
have been calculated over the total variable stars.}
\begin{tabular}{|l|c|c|c|}
\hline 
{\scriptsize Stars} & {\scriptsize G < 20 mag} & {\scriptsize Grvs < 17 mag} & {\scriptsize Grvs < 12 mag}\tabularnewline
\hline 
\hline 
{\scriptsize Single variable stars} & {\scriptsize 24.52\% } & {\scriptsize 25.79\% } & {\scriptsize 28.39\% }\tabularnewline
\hline 
{\scriptsize Variable stars in multiple systems} & {\scriptsize 75.48\% } & {\scriptsize 74.21\% } & {\scriptsize 71.61\% }\tabularnewline
\hline 
\emph{\scriptsize $\Rightarrow$ In binary systems} & {\scriptsize 55.74\% } & {\scriptsize 52.65\% } & {\scriptsize 38.49\% }\tabularnewline
\hline 
\emph{\scriptsize $\Rightarrow$ Others (ternary, etc.)} & {\scriptsize 19.73\% } & {\scriptsize 21.55\% } & {\scriptsize 33.12\% }\tabularnewline
\hline 
\hline 
{\scriptsize Total variable stars} & {\scriptsize 28,000,000} & {\scriptsize 19,000,000} & {\scriptsize 2,700,000}\tabularnewline
\hline 
{\scriptsize Individually observable} & {\scriptsize 21,500,000} & {\scriptsize 16,000,000} & {\scriptsize 2,000,000}\tabularnewline
\hline 
\hline 
{\scriptsize With planets} & {\scriptsize 2.09\% } & {\scriptsize 2.64\% } & {\scriptsize 2.09\% }\tabularnewline
\hline 
\end{tabular}
\par\end{centering}

\end{table}

\begin{figure}
\begin{centering}
\includegraphics[scale=0.5]{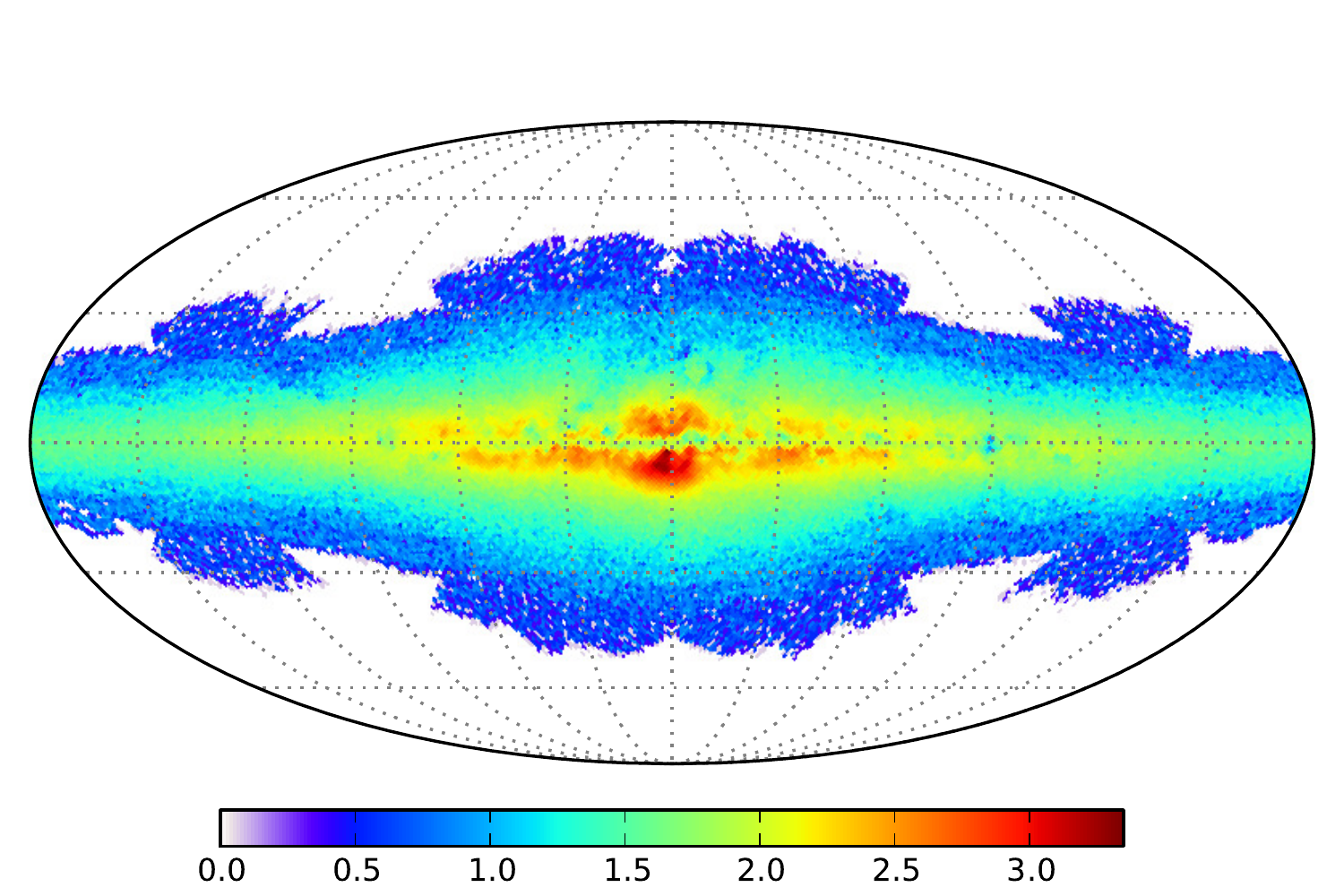}
\par\end{centering}

\caption{\label{fig:Sky-distribution-of-microlens}Sky distribution of microlenses
that could take place during the 5 years of the mission. Color scale
indicates the $\log_{10}$ of the number of microlenses per square
degree. }

\end{figure}

\begin{figure}
\begin{centering}
\includegraphics[scale=0.7]{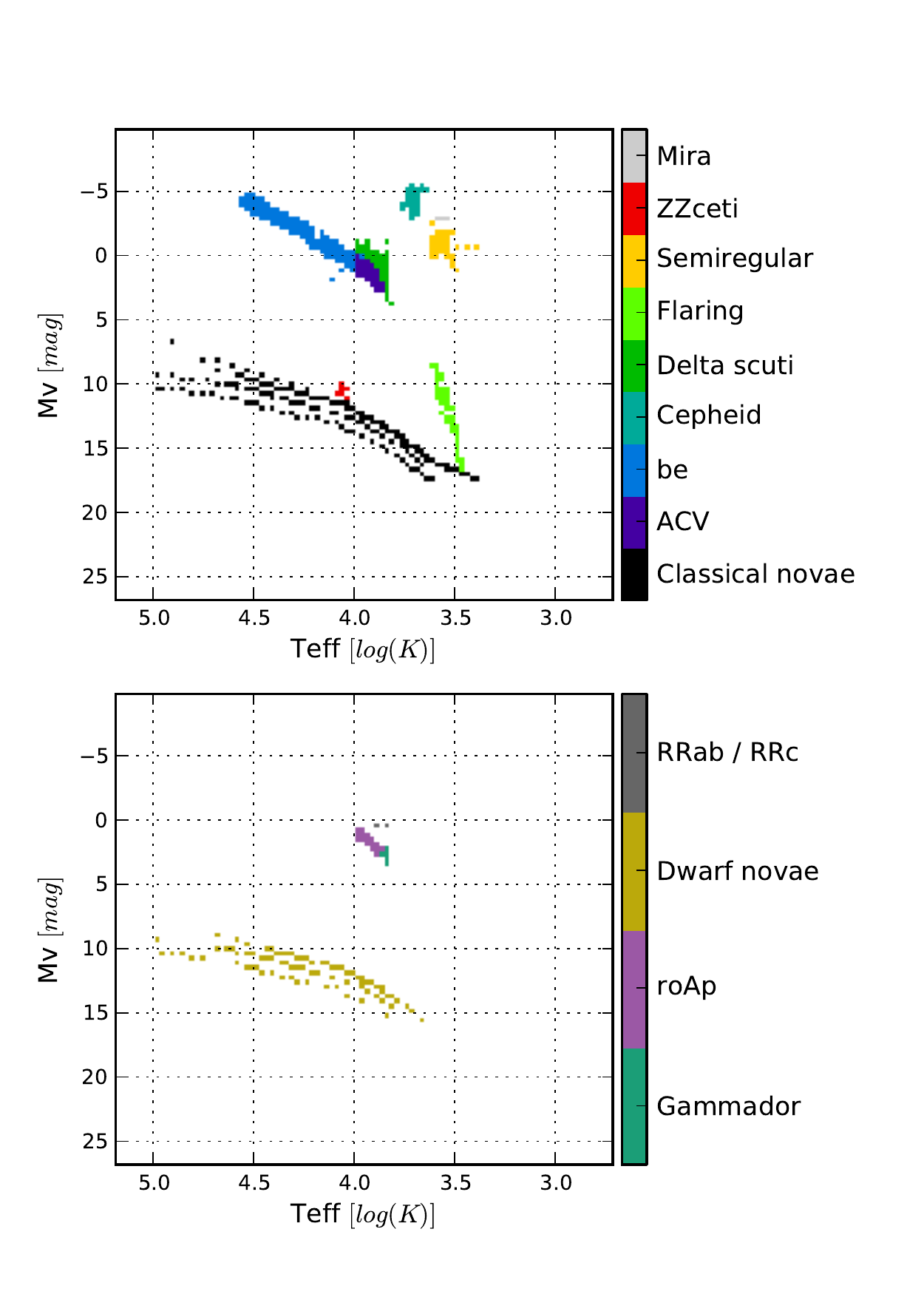}
\par\end{centering}

\caption{\label{fig:HR-diagram-split-var}HR diagram split by variability type. }

\end{figure}

By variability type, $\delta$ scuti are the most abundant representing
the 49\% of the variable stars, followed by semiregulars (42\%)
and microlenses (4.3\%) as seen in table \ref{tab:Stars-distribution-by-var}.
However, microlenses are highly related to denser regions of the galaxy
as shown in figure \ref{fig:Sky-distribution-of-microlens}, and they
can involve stars of any kind. The rest of variability types are strongly
related to different locations in the HR diagram (figure \ref{fig:HR-diagram-split-var}).

\begin{table}
\begin{centering}
\caption{\label{tab:Stars-distribution-by-var}Stars distribution by variability
type. }
\begin{tabular}{|c|c|c|c|}
\hline 
{\scriptsize Variability type} & {\scriptsize G < 20 mag} & {\scriptsize Grvs < 17 mag} & {\scriptsize Grvs < 12 mag}\tabularnewline
\hline 
\hline 
{\scriptsize ACV} & {\scriptsize 0.61\%} & {\scriptsize 0.52\%} & {\scriptsize 0.18\%}\tabularnewline
\hline 
{\scriptsize Flaring} & {\scriptsize 1.46\%} & {\scriptsize 0.49\%} & {\scriptsize 0.01\%}\tabularnewline
\hline 
{\scriptsize RRab} & {\scriptsize 0.37\%} & {\scriptsize 0.34\%} & {\scriptsize 0.02\%}\tabularnewline
\hline 
{\scriptsize RRc} & {\scriptsize 0.09\%} & {\scriptsize 0.09\%} & {\scriptsize 0.01\%}\tabularnewline
\hline 
{\scriptsize ZZceti} & {\scriptsize 0.12\%} & {\scriptsize <0.01\%} & {\scriptsize <0.01\%}\tabularnewline
\hline 
{\scriptsize Be} & {\scriptsize 2.15\%} & {\scriptsize 2.02\%} & {\scriptsize 0.87\%}\tabularnewline
\hline 
{\scriptsize Cepheids} & {\scriptsize 0.03\%} & {\scriptsize 0.04\%} & {\scriptsize 0.11\%}\tabularnewline
\hline 
{\scriptsize Classical novae} & {\scriptsize 0.05\%} & {\scriptsize 0.06\%} & {\scriptsize 0.19\%}\tabularnewline
\hline 
{\scriptsize $\delta$ scuti} & {\scriptsize 48.57\%} & {\scriptsize 41.01\%} & {\scriptsize 14.11\%}\tabularnewline
\hline 
{\scriptsize Dwarf novae} & {\scriptsize <0.01\%} & {\scriptsize <0.01\%} & {\scriptsize 0.00\%}\tabularnewline
\hline 
{\scriptsize Gammador} & {\scriptsize 0.09\%} & {\scriptsize 0.01\%} & {\scriptsize <0.01\%}\tabularnewline
\hline 
{\scriptsize Microlens} & {\scriptsize 4.27\%} & {\scriptsize 1.87\%} & {\scriptsize 0.91\%}\tabularnewline
\hline 
{\scriptsize Mira} & {\scriptsize 0.19\%} & {\scriptsize 0.24\%} & {\scriptsize 0.91\%}\tabularnewline
\hline 
{\scriptsize $\rho$ Ap} & {\scriptsize 0.05\%} & {\scriptsize 0.04\%} & {\scriptsize 0.01\%}\tabularnewline
\hline 
{\scriptsize Semiregular} & {\scriptsize 41.94\%} & {\scriptsize 53.27\%} & {\scriptsize 82.6\%}\tabularnewline
\hline 
\hline 
{\scriptsize Total} & {\scriptsize 21,500,000} & {\scriptsize 16,000,000} & {\scriptsize 2,000,000}\tabularnewline
\hline 
\end{tabular}
\par\end{centering}

\end{table}

\subsection{Binary stars}

As seen in section \ref{sub:Overview}, the model has generated 410,000,000
binary systems. Therefore about 820,000,000 stars have been generated
but it is important to remark that not all of them will be individually
observable by Gaia. Some systems may have components with magnitude
G fainter than 20 (although the integrated magnitude is brighter) and
others may be so close together that they cannot be resolved (although
they can be detected by other means).

The majority of the primary stars are from the main sequence (67\%),
being the most popular combination a double main sequence star system
(62\%). Sub-giants and giants as primary coupled with a main sequence
star are the second and third most probable systems (16\% and 14\%
 respectively). In general terms, the distribution is coherent with star formation and
evolution theories (e.g. supergiants are not accompanied by
white dwarfs), see table \ref{tab:Luminosity-class-combination}. 

\begin{table*}
\begin{centering}
\caption{\label{tab:Luminosity-class-combination}Binary stars classified depending
on the luminosity class combination (only binary systems whose integrated
magnitude is G<20). Values are in percentage. }
\begin{tabular}{|l|r|r|r|r|r|r|r||r|}
\cline{2-9} 
\multicolumn{1}{l|}{} & {\scriptsize \rotatebox[origin=c]{90}{Supergiant}} & {\scriptsize \rotatebox[origin=c]{90}{Bright Giant}} & {\scriptsize \rotatebox[origin=c]{90}{Giant}} & {\scriptsize \rotatebox[origin=c]{90}{Sub-giant}} & {\scriptsize \rotatebox[origin=c]{90}{Main sequence}} &  {\scriptsize \rotatebox[origin=c]{90}{White dwarf}} & {\scriptsize \rotatebox[origin=c]{90}{Others}} & {\scriptsize \rotatebox[origin=c]{90}{ Total per primary}}\tabularnewline
\hline 
Supergiant               &0.0000 &	0.0001	&0.0001	&0.0004	&0.0021	&0.0000	&0.0000	&0.0028\\
\hline
Bright giant             &0.0000	&0.0022	&0.0140	&0.0154	&0.1487	&0.0015	&0.0000	&0.1819\\\hline
Giant                    & 0.0000	&0.2933	&0.5933	&0.4997	&12.7477	&0.7229	&0.0072	&14.8641\\
\hline
Sub-giant                & 0.0001	&0.5135	&0.6916	&0.5429	&16.1521	&0.0000	&0.0100	&17.9101\\
\hline
Main sequence            & 0.0001	&0.4990	&1.1328	&0.0657	&63.0344	&1.8421	&0.0000	&66.5743\\
\hline
White dwarf              &0.0000	&0.0019&	0.0059	&0.0043	&0.4258	&0.0280	&0.0000&	0.4659\\
\hline
Others                   &0.0000	&0.0001	&0.0001	&0.0000	&0.0009	&0.0001	&0.0000&	0.0011\\
\hline
\hline
{ Total per secondary}&0.0002	&1.3101	&2.4378	&1.1283	&92.5118&	2.5946	&0.0172&
\multicolumn{1}{r}{}\\
\cline{1-8} 
\end{tabular}
\par\end{centering}

\end{table*}

\begin{figure}
\begin{centering}
\includegraphics[scale=0.6]{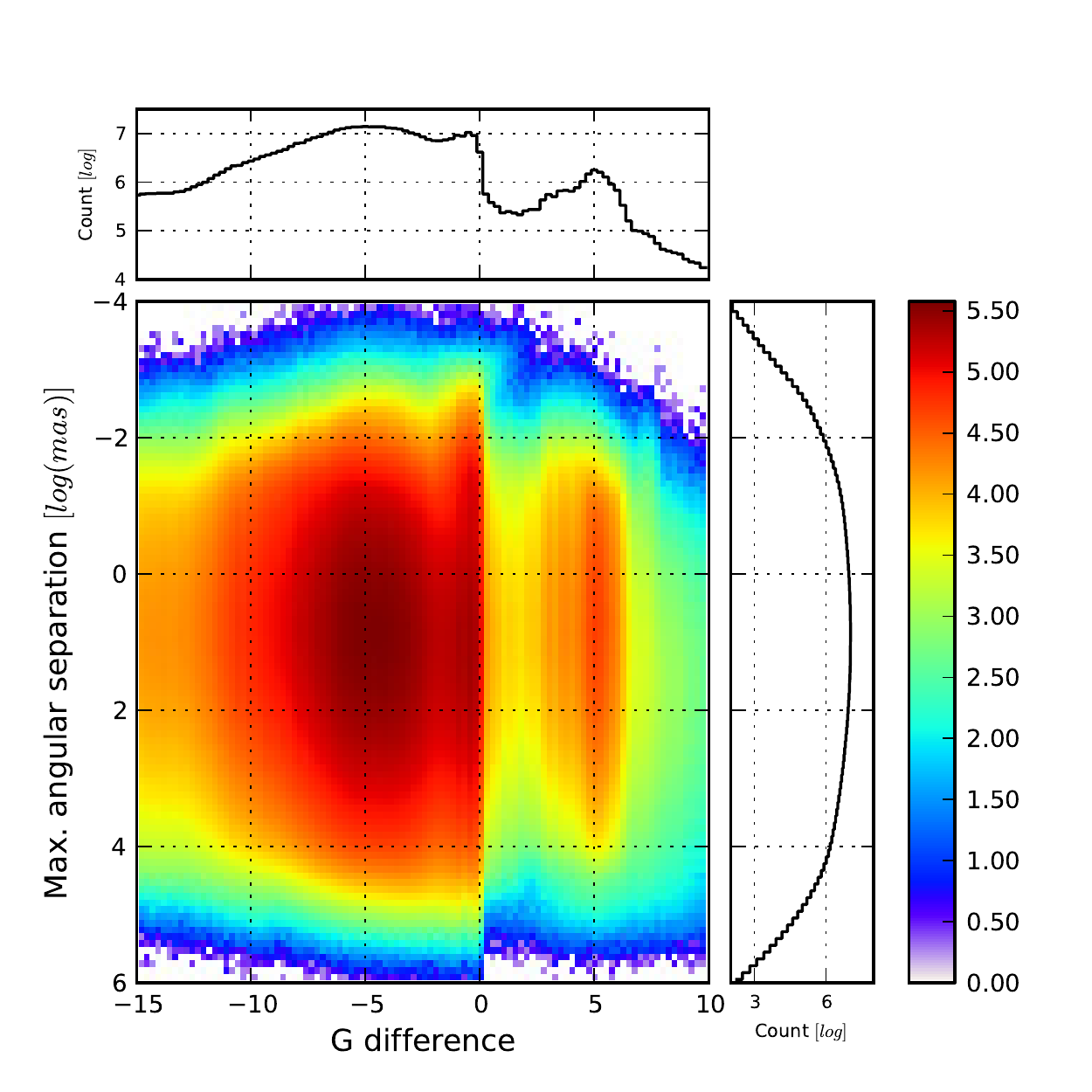}

\par\end{centering}

\caption{\label{fig:G-difference-and-angular-separation}G difference and angular
separation relation for binaries. Color scale indicates the $\log_{10}$
of the number of binaries per 0.25 difference in magnitude and $6\,\log\left(mas\right)$.}

\end{figure}

The magnitude difference versus
angular separation between components is shown 
in figure \ref{fig:G-difference-and-angular-separation}.
While main sequence pairs should produce only negative
magnitude differences, the presence of white dwarf primaries
with small red dwarf companions produce the asymmetrical shape
of the figure (recalling that "primary" here means the more massive).
To give a hint of the angular resolution capabilities of Gaia,
we can assume it at first step as nearly diffraction-limited 
and correctly sampled with pixel size $\approx 59$ mas.
Figure \ref{fig:G-difference-and-angular-separation} thus 
shows that a small fraction only of binaries with 
moderate magnitude differences will be resolved. 

\begin{figure}
\begin{centering}
\includegraphics[scale=0.6]{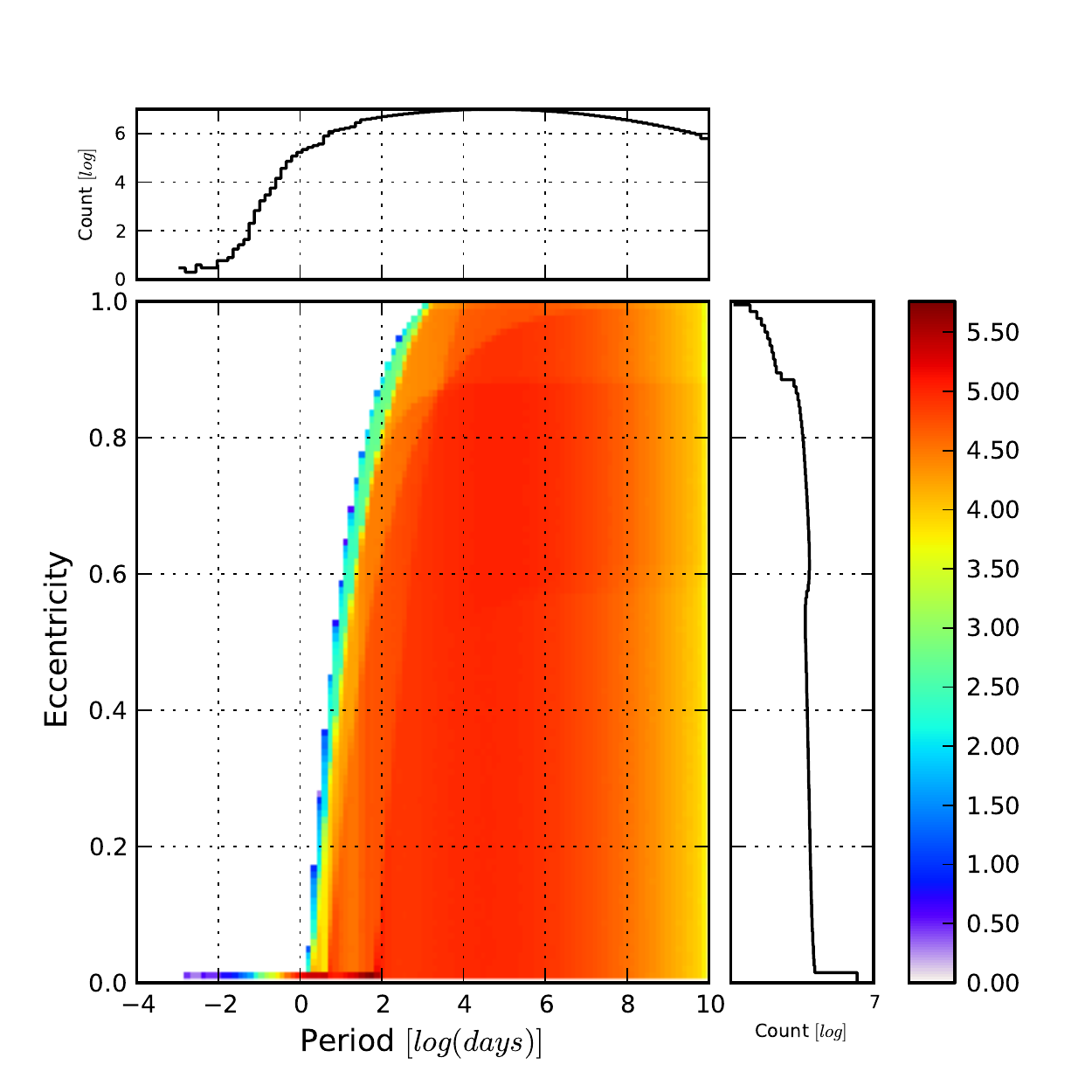}
\par\end{centering}

\caption{\label{fig:Eccentricity-Period} Period - Eccentricity relation for
binaries. Color scale indicates the $\log_{10}$ of the number of
binaries per 0.01 difference in eccentricity and $0.13\log\left(\mathrm{days}\right)$.}
\end{figure}

The mean separation of binary systems is $30$ AU and they present
a mean orbital period of about 250 years (figure \ref{fig:Eccentricity-Period}).
While only pairs with periods smaller than a decade may
have their orbit determined by Gaia, a significative fraction 
of binaries will be detected
through the astrometric ``acceleration'' of their motion.

\subsection{Stars with planets}

A total number of 34,000,000 planets have been generated and associated
to 27,500,000 single stars (\textasciitilde{}2.6\% over
total individually observable stars exposed in section \ref{sub:Overview}),
implying that 25\% of the stars have been generated with two planets
(table \ref{tab:Overview-of-stars-with-planets}). No exoplanets are
associated to multiple systems in this version of the model.

\begin{table}
\begin{centering}
\caption{\label{tab:Overview-of-stars-with-planets}Overview of stars with
planets}
\begin{tabular}{|l|c|c|c|}
\hline 
{\scriptsize Stars} & {\scriptsize G < 20 mag} & {\scriptsize Grvs < 17 mag} & {\scriptsize Grvs < 12 mag}\tabularnewline
\hline 
\hline 
{\scriptsize Total stars with planets} & {\scriptsize 27,500,000} & {\scriptsize 9,000,000} & {\scriptsize 182,000}\tabularnewline
\hline 
\hline 
{\scriptsize $\Rightarrow$}\emph{\scriptsize{} Stars with one planet} & \emph{\scriptsize 75.00\% } & \emph{\scriptsize 74.99\% } & \emph{\scriptsize 74.93\% }\tabularnewline
\hline 
{\scriptsize $\Rightarrow$}\emph{\scriptsize{} Stars with two planets} & \emph{\scriptsize 25.00\% } & \emph{\scriptsize 25.01\% } & \emph{\scriptsize 25.07\% }\tabularnewline
\hline 
\hline 
{\scriptsize Total number of planets} & {\scriptsize 34,000,000} & {\scriptsize 11,000,000} & {\scriptsize 228,000}\tabularnewline
\hline 
\end{tabular}
\par\end{centering}

\end{table}

The majority of star with planets belong to the main sequence (66\%),
followed by giants (17\%) and sub-giants (16.8\%) as shown in
table \ref{tab:Luminosity-class-of-stars-with-planets}. Only 8\%
of stars have a planet that produces eclipses.

\begin{table}
\begin{centering}
\caption{\label{tab:Luminosity-class-of-stars-with-planets}Luminosity class
of stars with planets}
\begin{tabular}{|c|c|c|c|}
\hline 
{\scriptsize Luminosity class} & {\scriptsize G < 20 mag} & {\scriptsize Grvs < 17 mag} & {\scriptsize Grvs < 12 mag}\tabularnewline
\hline 
\hline 
{\scriptsize Supergiant} & {\scriptsize <0.01\% } & {\scriptsize 0.01\% } & {\scriptsize 0.14\% }\tabularnewline
\hline 
{\scriptsize Bright giant} & {\scriptsize 0.18\% } & {\scriptsize 0.46\% } & {\scriptsize 2.50\% }\tabularnewline
\hline 
{\scriptsize Giant} & {\scriptsize 17.04\% } & {\scriptsize 34.05\% } & {\scriptsize 65.01\% }\tabularnewline
\hline 
{\scriptsize Sub-giant} & {\scriptsize 16.81\% } & {\scriptsize 14.07\% } & {\scriptsize 11.91\% }\tabularnewline
\hline 
{\scriptsize Main sequence} & {\scriptsize 65.72\% } & {\scriptsize 51.19\% } & {\scriptsize 20.36\% }\tabularnewline
\hline 
{\scriptsize Pre-main sequence} & {\scriptsize 0.20\% } & {\scriptsize 0.22\% } & {\scriptsize 0.08\% }\tabularnewline
\hline 
{\scriptsize White dwarf} & {\scriptsize 0.03\% } & {\scriptsize <0.01\% } & {\scriptsize <0.01\% }\tabularnewline
\hline 
{\scriptsize Others} & {\scriptsize <0.01\% } & {\scriptsize <0.01\% } & {\scriptsize <0.01\% }\tabularnewline
\hline 
\hline 
{\scriptsize Total} & {\scriptsize 27,500,000 } & {\scriptsize 9,000,000 } & {\scriptsize 182,000 }\tabularnewline
\hline 
\end{tabular}
\par\end{centering}

\end{table}

\begin{figure}
\begin{centering}

\includegraphics[scale=0.4]{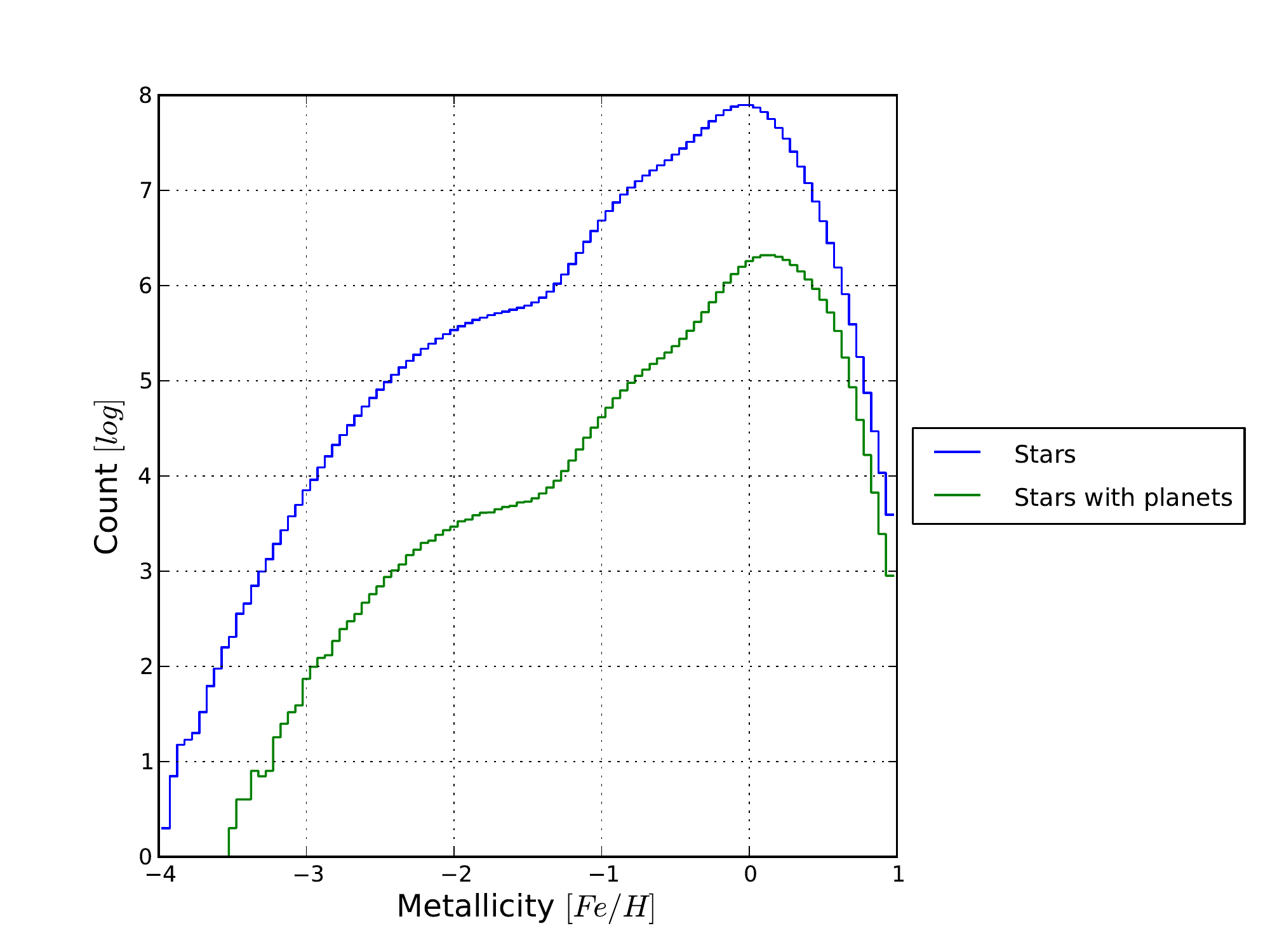}
\par\end{centering}

\caption{\label{fig:Metallicity-stars-with-planets}Metallicity distribution
of stars with planets. }

\end{figure}

On the other hand, stars with solar or higher metallicities
present a bigger probability of having a planet than stars poorer
in metals (figure \ref{fig:Metallicity-stars-with-planets}).

\begin{figure}
\begin{centering}

\includegraphics[scale=0.4]{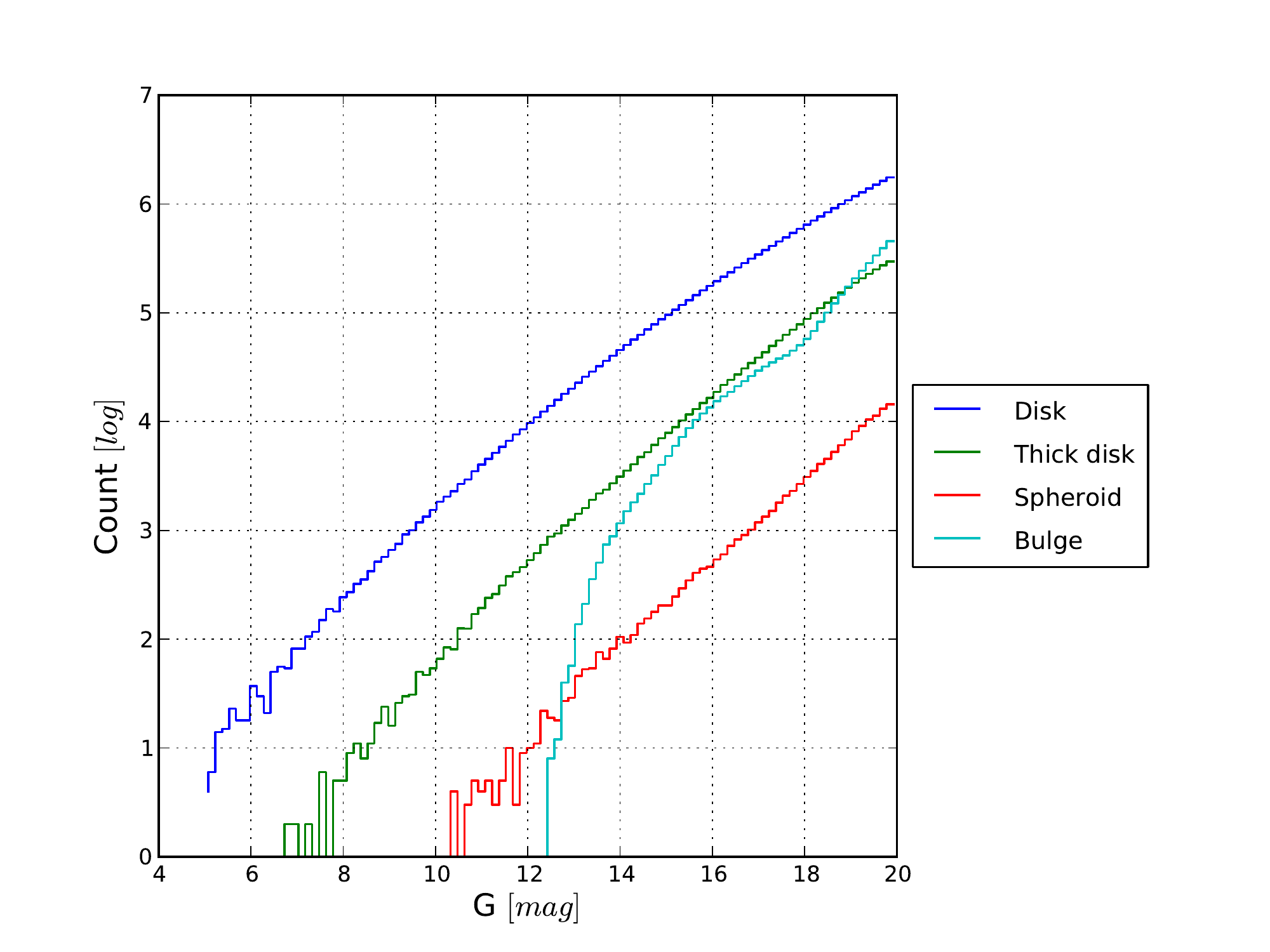}
\par\end{centering}

\caption{\label{fig:G-distribution-of-stars-planets-by-pop}G distribution
of stars having planets split by population.}

\end{figure}

Finally, 77\% of stars with planets belong to the thin disc population,
while 11\% are in the bulge, 11\% in the thick disc and 0.4\%
in the spheroid (figure \ref{fig:G-distribution-of-stars-planets-by-pop}).

\subsection{Extragalactic objects}

Apart from the stars presented in the previous sections, the model
generates 8,800,000 additional stars that belong to the Magellanic
Clouds. Again, the most abundant spectral type are G stars (46\%),
followed by K types (33\%) and A types (17\%). There
are no F type stars reachable by Gaia, because the magnitude cut at the cloud distance selects only the upper part of the HR diagram including massive stars on the blue side, and late type giants and supergiants on the red side.

\begin{table}
\begin{centering}
\caption{\label{tab:Overview-of-extragalactic}Overview of extragalactic objects}
\begin{tabular}{|l|c|c|c|}
\hline 
{\scriptsize Stars} & {\scriptsize G < 20 mag} & {\scriptsize Grvs < 17 mag} & {\scriptsize Grvs < 12 mag}\tabularnewline
\hline 
\hline 
{\scriptsize Stars in LMC} & {\scriptsize 7,550,000} & {\scriptsize 1,039,000 } & {\scriptsize 5,600 }\tabularnewline
\hline 
{\scriptsize Stars in SMC} & {\scriptsize 1,250,000} & {\scriptsize 161,000} & {\scriptsize 950}\tabularnewline
\hline 
{\scriptsize Unresolved galaxies} & {\scriptsize 38,000,000} & {\scriptsize 3,000,000} & {\scriptsize 4,320}\tabularnewline
\hline 
{\scriptsize QSO} & {\scriptsize 1,000,000} & {\scriptsize 5,200} & {\scriptsize 11}\tabularnewline
\hline 
{\scriptsize Supernovae} & {\scriptsize 50,000} & {\scriptsize -} & {\scriptsize -}\tabularnewline
\hline 
\end{tabular}
\par\end{centering}

\end{table}

\begin{table}
\begin{centering}
\caption{\label{tab:LMC-SMC-Stars}Spectral types of stars from LMC/SMC}
\begin{tabular}{|c|c|c|c|}
\hline 
{\scriptsize Spectral type} & {\scriptsize G < 20 mag} & {\scriptsize Grvs < 17 mag} & {\scriptsize Grvs < 12 mag}\tabularnewline
\hline 
\hline 
{\scriptsize O} & {\scriptsize 0.25\% } & {\scriptsize 0.17\% } & {\scriptsize 0.39\% }\tabularnewline
\hline 
{\scriptsize B} & {\scriptsize 3.24\% } & {\scriptsize 3.40\% } & {\scriptsize 1.85\% }\tabularnewline
\hline 
{\scriptsize A} & {\scriptsize 17.20\% } & {\scriptsize 5.01\% } & {\scriptsize 4.83\% }\tabularnewline
\hline 
{\scriptsize F} & {\scriptsize 0.00\% } & {\scriptsize 0.00\% } & {\scriptsize 0.00\% }\tabularnewline
\hline 
{\scriptsize G} & {\scriptsize 45.98\% } & {\scriptsize 23.16\% } & {\scriptsize 55.19\% }\tabularnewline
\hline 
{\scriptsize K} & {\scriptsize 32.62\% } & {\scriptsize 64.82\% } & {\scriptsize 35.33\% }\tabularnewline
\hline 
{\scriptsize M} & {\scriptsize 0.71\% } & {\scriptsize 3.44\% } & {\scriptsize 2.41\% }\tabularnewline
\hline 
{\scriptsize Total} & {\scriptsize 8,800,000 } & {\scriptsize 1,200,000 } & {\scriptsize 6,600 }\tabularnewline
\hline 
\end{tabular}
\par\end{centering}

\end{table}

Regarding unresolved galaxies, 38,000,000 have been generated. However due to the 
on board thresholding algorithm optimized for point sources, a significant fraction of these galaxies will
not have their data transfered to Earth. 
Gaia will be able to measure radial velocities for only 8\% and
metal abundances for 0.01\% of them. The most frequent galaxy type
are spirals (58\% considering Sa, Sb, Sc, Sd, Sbc), followed by
irregulars (25\%) and ellipticals (13\% adding E2 and E-S0)
as seen in table \ref{tab:Galaxies-by-type}.

\begin{table}
\begin{centering}
\caption{\label{tab:Galaxies-by-type}Galaxies by type. Percentages calculated
over total galaxies (table \ref{tab:Overview-of-extragalactic}).}
\begin{tabular}{|l|c|c|c|}
\hline 
{\scriptsize Galaxies} & {\scriptsize G < 20 mag} & {\scriptsize Grvs < 17 mag} & {\scriptsize Grvs < 12 mag}\tabularnewline
\hline 
\hline 
{\scriptsize E2} & {\scriptsize 6.36\%} & {\scriptsize 10.24\%} & {\scriptsize 10.51\%}\tabularnewline
\hline 
{\scriptsize E-S0} & {\scriptsize 7.03\%} & {\scriptsize 11.67\%} & {\scriptsize 12.85\%}\tabularnewline
\hline 
{\scriptsize Sa} & {\scriptsize 7.51\%} & {\scriptsize 10.55\%} & {\scriptsize 10.05\%}\tabularnewline
\hline 
{\scriptsize Sb} & {\scriptsize 9.21\%} & {\scriptsize 12.39\% } & {\scriptsize 13.70\% }\tabularnewline
\hline 
{\scriptsize Sc } & {\scriptsize 10.21\% } & {\scriptsize 8.50\% } & {\scriptsize 8.08\% }\tabularnewline
\hline 
{\scriptsize Sd } & {\scriptsize 22.08\% } & {\scriptsize 17.08\% } & {\scriptsize 15.00\% }\tabularnewline
\hline 
{\scriptsize Sbc } & {\scriptsize 9.21\% } & {\scriptsize 12.35\% } & {\scriptsize 13.73\% }\tabularnewline
\hline 
{\scriptsize Im } & {\scriptsize 24.77\% } & {\scriptsize 15.46\% } & {\scriptsize 14.44\% }\tabularnewline
\hline 
{\scriptsize QSFG } & {\scriptsize 3.61\% } & {\scriptsize 1.76\% } & {\scriptsize 1.64\% }\tabularnewline
\hline 
\end{tabular}
\par\end{centering}

\end{table}

Additionally, 1,000,000 quasars have been generated and the model
predicts 50,000 supernovas that will occur during the 5 years of mission,
being 77\% of them from Ia type (table \ref{tab:Supernovae-types}).

\begin{table}
\begin{centering}
\caption{\label{tab:Supernovae-types}Supernovae types. Percentages calculated
over 50,000 supernovas.}
\begin{tabular}{|c|c|}
\hline 
{\scriptsize Supernovae types} & \multicolumn{1}{c|}{}\tabularnewline
\hline 
\hline 
{\scriptsize Ia} & {\scriptsize 76.74\%}\tabularnewline
\hline 
{\scriptsize Ib/c} & {\scriptsize 7.36\%}\tabularnewline
\hline 
{\scriptsize II-L} & {\scriptsize 14.24\%}\tabularnewline
\hline 
{\scriptsize II-P} & {\scriptsize 1.65\%}\tabularnewline
\hline 
\end{tabular}
\par\end{centering}

\end{table}

\begin{figure}
\begin{centering}
\includegraphics[scale=0.55]{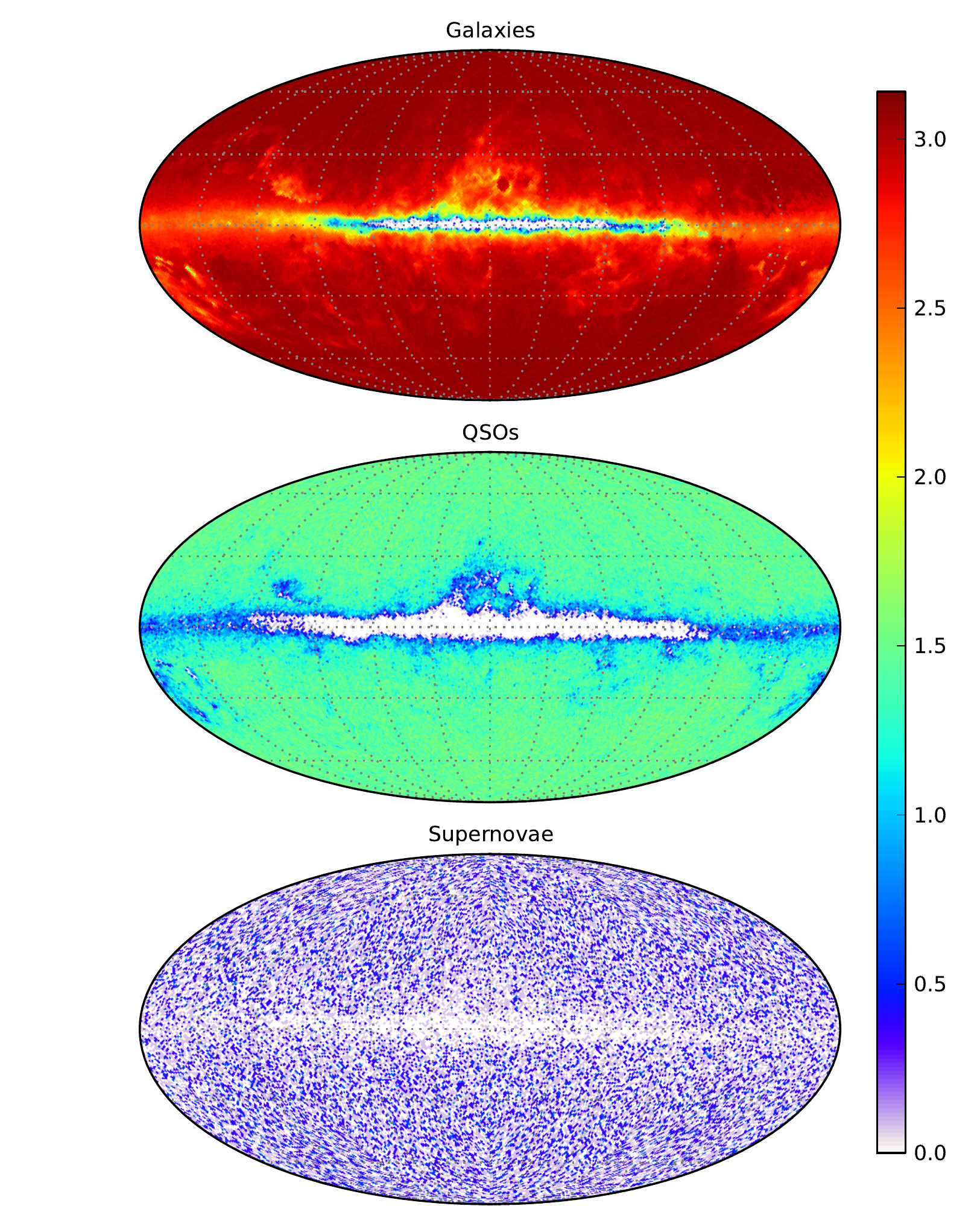}
\par\end{centering}

\caption{\label{fig:Total-sky-distribution-extragalactic}Total sky distribution
of unresolved galaxies, quasars and supernovae. Color scale indicates
the $\log_{10}$ of the number of objects per square degree.}
\end{figure}

The sky distribution of these three types of objects is shown in figure
\ref{fig:Total-sky-distribution-extragalactic}.

\begin{figure}
\begin{centering}
\includegraphics[scale=0.65]{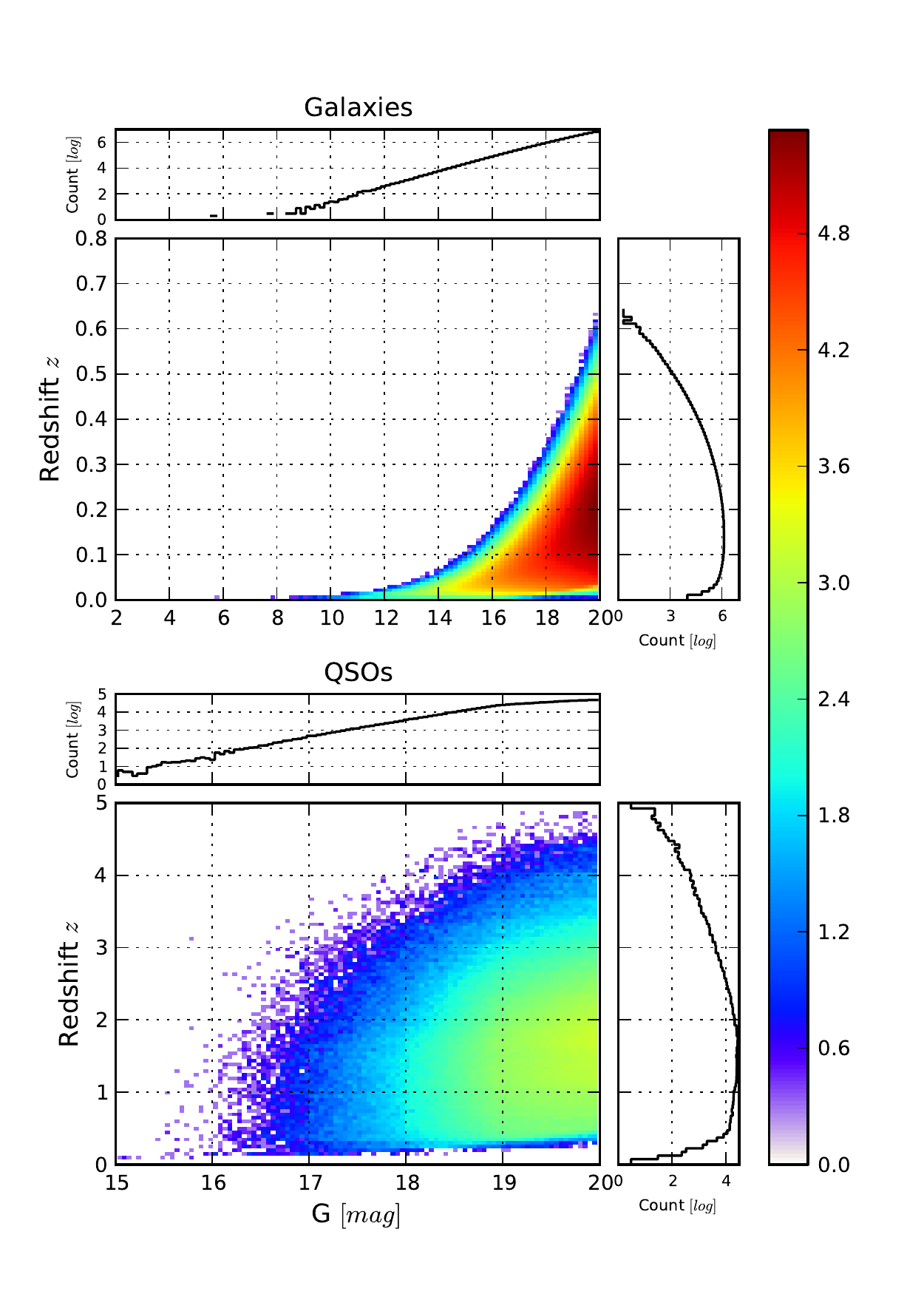}

\par\end{centering}

\caption{\label{fig:Redshift-and-G}Redshift and G relation for galaxies and
QSOs. Color scale indicates the $\log_{10}$ of the number of objects
per 0.05 mag and 0.05 redshift difference.}
\end{figure}

Regarding redshifts, galaxies go up to $z\sim0.75$, while the most distant
quasars are at $z\sim5$. Magnitudes also present a different pattern
depending on the type of object as shown in figure \ref{fig:Redshift-and-G}.

\section{Conclusions}

Gaia will be able to provide a much more precise and complete view
of the Galaxy than its predecessor Hipparcos, representing a large
increase in the total number of stars (table \ref{tab:Hipparcos-Gaia})
and the solar neighbourhood (table \ref{tab:Hipparcos-Gaia-nearby-stars}).

\begin{table}
\begin{centering}
\caption{\label{tab:Hipparcos-Gaia}Comparison of Hipparcos and Gaia characteristics
and the predicted number of stars, variables and binaries presented
in this study.\protect \\
\emph{\scriptsize {*} Includes stars which may not be resolved due
to its angular separation and Gaia resolution power.}}
\begin{tabular}{|c|c|c|}
\hline 
 & {\scriptsize Hipparcos} & {\scriptsize Gaia}\tabularnewline
\hline 
\hline 
{\scriptsize Number of stars } & {\scriptsize 118,218} & {\scriptsize $1,100,000,000{}^{*}$}\tabularnewline
\hline 
{\scriptsize Mean sky density (per square degree)} & {\scriptsize \textasciitilde{} 3} & {\scriptsize \textasciitilde{} 30.425}\tabularnewline
\hline 
{\scriptsize Limiting magnitude } & {\scriptsize V \textasciitilde{} 12.4 mag} & {\scriptsize V \textasciitilde{} 20 \textendash{} 25 mag}\tabularnewline
\hline 
{\scriptsize Median astrometric precision} & {\scriptsize 0.97 mas (V<9)} & {\scriptsize \textasciitilde{} 10$\mu$ as (V=15)}\tabularnewline
\hline 
{\scriptsize Possibly variable } & {\scriptsize 11,597} & {\scriptsize $21,500,000^{*}$}\tabularnewline
\hline 
{\scriptsize Suspected double systems } & {\scriptsize 23,882} & {\scriptsize $410,000,000^{*}$}\tabularnewline
\hline 
\end{tabular}
\par\end{centering}

\end{table}

\begin{table}
\begin{centering}
\caption{\label{tab:Hipparcos-Gaia-nearby-stars}Number of nearby stars (within 25 pc)
detected by Hipparcos compared with the predicted number of stars in the universe
model}
\begin{tabular}{|c|c|c|c|}
\hline 
\multicolumn{2}{|c|}{{\scriptsize Distance (pc)}} & \multicolumn{2}{c|}{{\scriptsize Stars}}\tabularnewline
\hline 
{\scriptsize Min.} & {\scriptsize Max.} & {\scriptsize Hipparcos (V < 9)} & {\scriptsize Gaia (G < 20)}\tabularnewline
\hline 
\hline 
{\scriptsize 20} & {\scriptsize 25} & {\scriptsize 1,126} & {\scriptsize 4,563}\tabularnewline
\hline 
{\scriptsize 10} & {\scriptsize 20} & {\scriptsize 1,552} & {\scriptsize 1,784}\tabularnewline
\hline 
{\scriptsize 5} & {\scriptsize 10} & {\scriptsize 257} & {\scriptsize 927}\tabularnewline
\hline 
{\scriptsize 0} & {\scriptsize 5} & {\scriptsize 61} & {\scriptsize 706}\tabularnewline
\hline 
{\scriptsize 0} & {\scriptsize 25} & {\scriptsize 2,996} & {\scriptsize 7,980}\tabularnewline
\hline 
\end{tabular}
\par\end{centering}

\end{table}

The Gaia Universe Model, and other population synthesis models in
general, can be useful tools for survey preparation. In this particular
case, it has been possible to obtain a general idea of the numbers,
percentages and distribution of different objects and characteristics
of the environment that Gaia can potentially observe and measure. 

Additionally, the analysis of the snapshot has facilitated the detection
of several aspects to be improved. Therefore,
it has been a quite fruitful quality assurance process from a scientific
point of view.

Looking forward, the next reasonable step would be to reproduce the
same analysis but taking into consideration the instrument specifications
and the available error models. By convolving it with the idealized
universe model presented in this paper, it will be possible to evaluate
the impact of the instrumental effects on the actual composition of
the Gaia catalogue. In the mean time the simulated catalogue 
presented in this paper will be publicly available through the ESA Gaia 
Portal at \url{http://www.rssd.esa.int/gaia/}.

\begin{acknowledgements}
The development of the Universe Model and of the GOG simulator used here have been possible thanks to contributions from many people. We warmly acknowledge the contributions from Mary Kontizas, Laurent Eyer, Guillaume Debouzy, Oscar Mart\'{\i}nez, Raul Borrachero, Jordi Peralta, and Claire Dollet.

This work has been possible thanks to the European Science Foundation
(ESF) and its funding for the activity entitled 'Gaia Research for
European Astronomy Training', as well as the infrastructure used for
the computation in the Barcelona Supercomputing Center (MareNostrum).

This work was supported by the MICINN (Spanish Ministry of Science and Innovation) - 
FEDER through grant AYA2009-14648-C02-01 and CONSOLIDER CSD2007-00050. 

\end{acknowledgements}
\emph{\scriptsize NOTE: Internal referred documents from the Gaia mission can be found at \url{http://www.rssd.esa.int/index.php?project=GAIA&page=Library_Livelink}}{\scriptsize \par}

\bibliographystyle{aa}
\bibliography{Robin2012}

\end{document}